\documentclass[useAMS,usenatbib]{mn2e}

\usepackage{natbib}				
\usepackage{epsf} 
\usepackage{graphicx}
\usepackage{algorithm2e}
\usepackage{enumerate}
\usepackage{multirow}
\usepackage{enumitem}
\newcommand{\cbf}[1]{ \textbf{\textit{#1}}}
\usepackage{array}
\usepackage{amsmath}
\usepackage[T1]{fontenc}
\usepackage{aecompl}
\usepackage{url}
\usepackage{makecell}

\usepackage[usenames, dvipsnames]{color}

\newcolumntype{C}[1]{>{\centering}m{#1}}

\title[\texttt{ASTErIsM} ]
{\texttt{ASTErIsM} - Application of topometric  clustering algorithms in automatic galaxy detection and   classification}

\author[A. Tramacere et al.]{A. Tramacere$^{1}$\thanks{E-mail: andrea.tramacere@unige.ch (AT); andrea.tramacere@gmail.com (AT)},
	D. Paraficz$^2$,
	P. Dubath $^1$,
	J.-P. Kneib $^2$,
	F. Courbin $^2$\\
$^{1}$ISDC, Department of Astronomy University of Geneva, 16, 1290, Versoix, Switzerland\\
$^{2}$Laboratoire d'Astrophysique Ecole Polytechnique Federale de Lausanne  (EPFL) Observatoire de Sauverny CH-1290 Versoix\\}

\begin{document}


\pagerange{\pageref{firstpage}--\pageref{lastpage}} \pubyear{2002}

\maketitle

\label{firstpage}

\begin{abstract}
We present a study on galaxy detection and shape classification using  topometric clustering algorithms. We  first use the DBSCAN algorithm to extract, from CCD frames, groups of adjacent pixels with significant fluxes and we then apply the DENCLUE algorithm to separate the contributions of overlapping sources. The DENCLUE  separation is based on the localization of pattern of local maxima, through an iterative algorithm which associates each pixel to the closest local maximum.

Our main classification goal is to take apart elliptical from spiral galaxies. We introduce new sets of features derived from the computation of geometrical invariant moments of the pixel group shape and from the statistics of the spatial distribution of the DENCLUE local maxima patterns.  Ellipticals are characterized by a single group of local maxima, related  to the galaxy core, while spiral galaxies have additional ones related to segments of spiral arms. We use two different supervised ensemble classification algorithms, Random Forest, and Gradient Boosting. Using a sample of $\simeq 24000$ galaxies taken from the Galaxy Zoo 2 main sample with spectroscopic redshifts, and we test our classification against the Galaxy Zoo 2 catalog.

We find that features extracted from our pipeline give on average an accuracy of $\simeq 93\%$, when testing on a test set with a size of $20\%$ of our full data set, with features 
deriving from the angular distribution of density attractor ranking at the top of the discrimination power.

\end{abstract}


\section{Introduction}


Morphology is one of the main characteristics of galaxies, as physical process happening during life time of galaxies strongly determine their shape. Therefore any theory of galaxy formation and evolution needs to closely explain the observational distribution of morphological classes \citep{Dressler1980, Bamford2009, Roberts1994}. Accurate information of galaxy types gives insight also well beyond galaxy research,  testing cosmological models by studying large scale structure with ETG clustering \citep{Naab:2007}, dark mater probe by strong gravitational lensing \citep{Koopmans2004, Treu:2002}

The key challenge of all this research is accurate and efficient classification of big number of galaxies. Traditional method of morphological classification classifies  galaxies according to Hubble's scheme  \citep{Sandage:1961}. This  system classifies the galaxy morphologies into elliptical, lenticular, spiral, and irregular galaxies. However, due to the impressive amount of photometric data produced
by large galaxy survey the size and quality modern data sets led to refinements in the classification \citep{Kormendy1996,Cappellari2011,vanderwel2007,Kartaltepe2015}

One possible classification methods is given by citizen science projects.
Excellent example of collaborative work on visual galaxy morphology classification is the Galaxy Zoo project that involved more than 100,000 volunteers to determine a galaxy class for about 900,000 (Galaxy Zoo 1) and 304,122(Galaxy Zoo 2) galaxies \citep{Lintott2011,Willett:2013ea}

On the other side, larger number of galaxies available in the next generation of all-sky survey missions (Euclid, LSST, KIDS etc) makes such a human-eye analysis unmanageable. Thus, to classify large numbers of galaxies into early and late types it is compelling to use instead automated morphological classification methods. 

 Several methods have been used to tackle this challenge, i.e. neural networks  \citep{Naim1995, Lahav:1996,Goderya:2002, Banerji:2010iq,Dieleman2015} , decision trees 
\citep{Owens1996}   and ensembles of classifiers \citep{Bazeil:2001}.

However, because visual inspection requires significant spatial resolution, it  is limited in  galaxy sample size and it is burden with possibility of missing rare and interesting objects due to lack of scientific knowledge of volunteers. Moreover, significantly larger number of galaxies available in the next generation of all-sky survey missions (Euclid, LSST, KIDS etc) makes such a human-eye analysis  unfeasible.


In this paper we present an  automated approach,  based on a novel combination  
of two topometric clustering  algorithms: the DBSCAN \citep{Ester96DBScan}, 
and the DENCLUE \citep{Hinneburg:1998wo,Hinneburg:2007tq}.
The DBSCAN algorithm has been already successfully applied to the detection 
of sources in $\gamma-$ray photons lists \citep{Tramacere:2013it,Carlson2013} and  to identify structure in external galaxies \citep{Rudick:2009}, while 
the DENCLUE, to our knowledge, has never been used, so far, in treatment of astronomical images.
In particular we  have noted that the DENCLUE algorithm, is effective both in the deblending of confused sources, and in the tracking  of  spiral arms. 
We have used these algorithms to  develop  a python  package: \texttt{ASTErIsM}	 ( python AStronomical Tools for clustering-based dEtectIon and Morphometry). This software is used  both to detect the  sources in CCD images, and to extract  features relevant to the morphological classification.

\begin{figure}
	\centering
	\begin{tabular}{l} 
		\includegraphics[width=8cm]{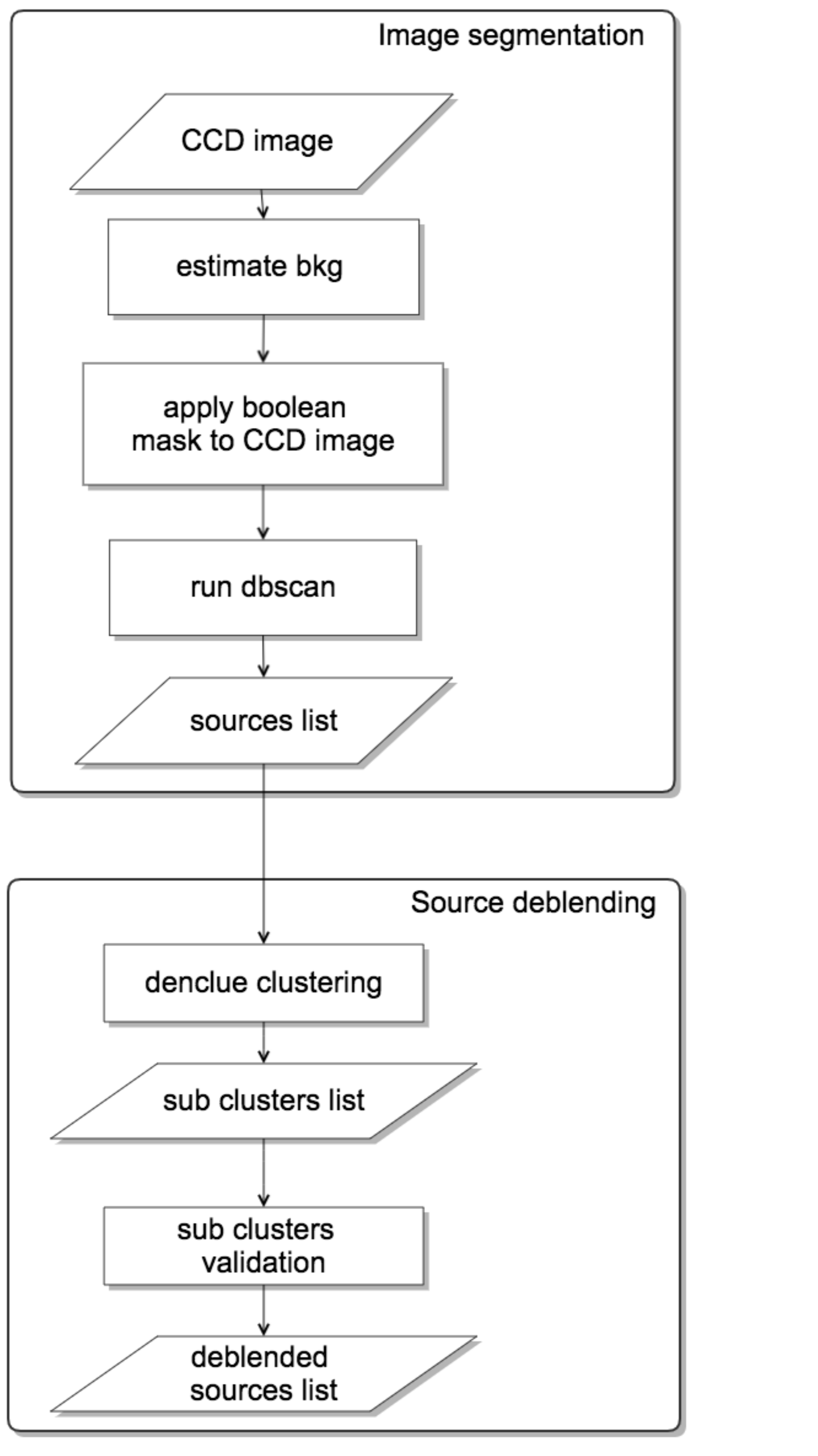}\\
		
	\end{tabular}
	\caption{Flow chart diagram for the detection process. }
	\label{fig:detection_flwchart}
\end{figure}

The paper is organized as follows: In Sec. \ref{sec:detection}, we describe the detection process, 
and in particular how we modified DBSCAN and DENCLUE algorithms to work 
on CCD images. In Sec. \ref{sec:denclue_spir_arms}, we describe how the DENCLUE method can be used 
to track spiral arms. In Sec. \ref{sec:pipeline_general} we present a general 
view of the \texttt{ASTErIsM} pipeline for automatic detection and morphological classification, 
the sample used in our paper,  the feature 
extractions process, and their statistical characterization.
In Sec. \ref{sec:trainign_sets} we describe the setup of the training sets.
In Sec. \ref{sec:superv_class} we describe the algorithms used for our supervised 
classification (Random Forest \citep{Breiman:2003} and  Gradient Boosting \citep{Friedman:2001ic}), and 
the metrics used for the classification.
In Sec. \ref{sec:class_strategy} we describe the strategy of our machine learning classification 
pipeline, and in Sec. \ref{sec:class_performance} we present the results of the classification 
together with a comparison to other similar works. In Sec. 
 \ref{sec:conclusions}, we present our final conclusions and future developments.
 The code will be available at \url{https://github.com/andreatramacere/asterism}

\begin{figure*}
	\centering
	\begin{tabular}{lll} 
		\includegraphics[width=5cm]{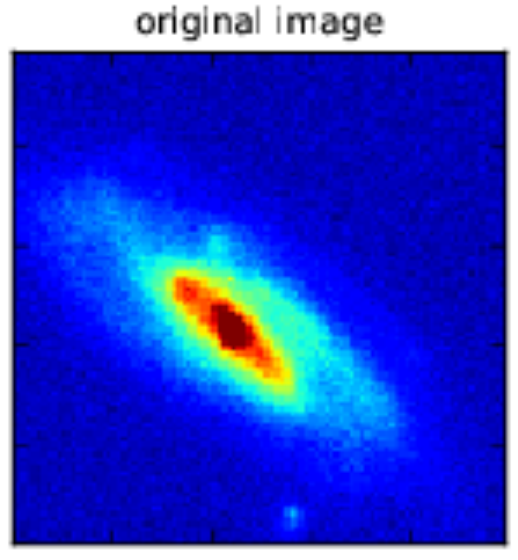}&
		\includegraphics[width=5cm]{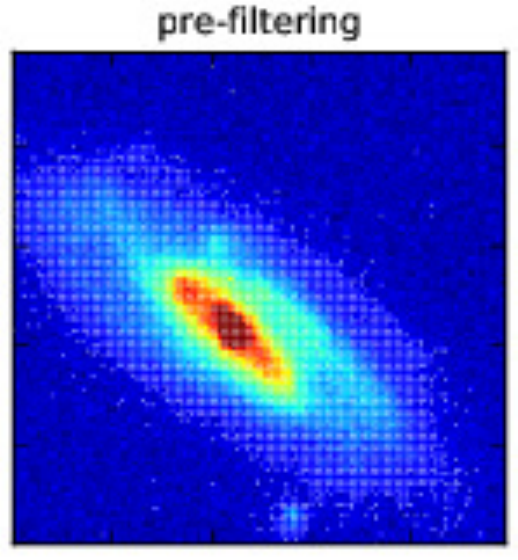}&
		\includegraphics[width=5cm]{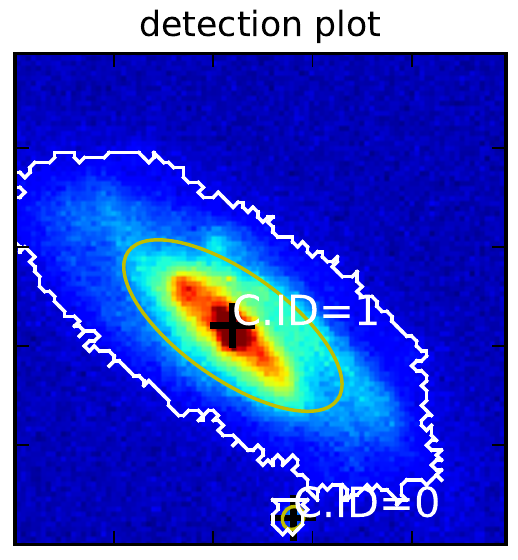}\\
	\end{tabular} 
	\caption{Application of the DBSCAN algorithm to source detection (see Sec. \ref{sec:source  detection}). The input  image is a {\it gri} summed bands
		image  cutout centered on  the object with  DR8OBJID=1237667549806657543 from the Galaxy Zoo 2 Main sample with spectroscopic redshifts.
		{\it Left panel:} the original image. {\it Central panel:} pixels selected (white dots), 
		with flux values above the background threshold. {\it Right panel:} the tow sources detected by the 
		DBSCAN algorithm, with $N_{bkg	}=3.9$, $K=1.5$ and $\varepsilon=1.0$. The withe line shows the source contour, the black crosses show the source centroid, and the yellow ellipses represent the containment ellipsoid.}
	\label{fig:fig_dbs_detection}
\end{figure*}

\section[]{Application of density-based  clustering methods to detect sources in CCD images }
\label{sec:detection}
The "density based spatial clustering of applications with noise" algorithm (DBSCAN) 
\citep{Ester96DBScan,Zaki:2014}, is a topometric density-based clustering method that  uses local density of  points to find clusters, in data sets that are affected by background noise.
Let $N_{\varepsilon}(p_i)$ be the  set of points contained within the N-dimensional sphere of radius $\varepsilon$ centered on  $p_i$, and $|N_{\varepsilon}(p_i)|$ the number of contained points, i.e. the estimator  of the local density, and $K$ a threshold value. Clusters are built according to the 
local density around  each point $p_i$. A  point is  classified according to the local density  defined as:

\begin{itemize}[leftmargin=0.5cm]
	\item \cbf{core point}: if   $|N_{\varepsilon}(p_i)|\geq K$ .
	\item \cbf{border point}: if  $|N_{\varepsilon}(p_i)|< K$, but
	at least one core point belongs to $N_{\varepsilon}(p_i)$. 
	\item \cbf{noise point}, if both  the conditions above are not satisfied.
\end{itemize}
Points are classified according to their inter-connection as: 
\begin{itemize}[leftmargin=0.5cm]
\item \cbf{directly density reachable}:  a point $p_j$  is defined \textit{ directly density reachable} from a point $p_k$, if $p_j \in N_{\varepsilon}(p_k) $ and $p_k$ is a \textit{core} point.
\item \cbf{density reachable}: a point $p_j$  is defined \textit{ density reachable} from a point $p_k$, if exists a chain of  \textit{ directly density reachable} points connecting, 
 $p_j$  to  $p_k$. 
\item \cbf{density connected}: two points $p_j$, $p_k$  are defined \textit{density connected}  if exits a \textit{core} point $p_l$ such that both  $p_j$ and $p_k$, are  \textit{density reachable} from $p_l$.
\end{itemize}
 The DBSCAN builds the cluster by progressively connecting \textit{density connected} 
 points to each set of \textit{core} points found in the set.  Thanks to its embedded capability  to distinguish background noise  (even when the background is not uniform),  it has been successfully used to detect  sources in $\gamma-$ray  photon lists \citep{Tramacere:2013it}, 
 or to identify structures in N-body simulations of galaxy clusters \citep{Rudick:2009}. 
For a  detailed description of the application of the DBSCAN  to $\gamma-$ray  photon lists, we address the reader to \cite{Tramacere:2013it}.
In general, a photon detection event will be characterized by position 
in detector/sky coordinates, and further possible features (energy, arrival time, etc...).
In the case of photon lists (as in $\gamma-$ray astronomy),  the detector/sky coordinates of each event can be recorded 
and the DBSCAN algorithm can be applied  to look for density-based  clusters, where a cluster is an  
astronomical source.  Events non belonging to any source (clusters), are assigned to background (noise).
In the case of CCD images, a DBSCAN suitable representation of the data is less intuitive. Indeed, detected photons  are not stored as single events, being    integrated and  positionally binned in the CCD matrix itself.
Since  the pixels coordinates have a uniform spatial distribution, we can not use the original estimator 
of local density $|N_{\varepsilon}(p_i)|$ to find density based clusters.
To overcome this limitation we have modified the DBSCAN algorithm basing on  the idea to use the photon counts/flux recorded in each pixel of the CCD as a new  estimator of the local density.

\subsection{Image segmenation: DBSCAN}
\label{sec:source detection}
In our modified version of  the DBSCAN algorithm  we have changed the procedure for the estimation of the local density as follows:
\begin{itemize}[leftmargin=0.5cm]
	\item We iterate through each pixel $p_{k,l}$, where $k$ refer to the $k_{th}$ row and $l$ to the $l_{th}$ column of the CCD matrix
	\item Let  $B_{\varepsilon}(k,l)$ be the set of pixels contained in the box  centered on the pixel  $p_{k,l}$, and enclosing the pixels with columns $k\pm\varepsilon$ and row  $l\pm\varepsilon$ 
	\item  We evaluate the local flux  $M_\varepsilon(k,l)$ as total flux  collected in $B_{\varepsilon}(k,l)$: 
	\begin{equation}
	 M_\varepsilon (k,l)=  \sum_{ (i,j) \in B_{\varepsilon}(k,l) }  I(p_{i,j})
	\end{equation}
\end{itemize}
 The quantity  $M_\varepsilon(k,l)$ is our estimator for the local density. With this choice the classification in {\it core, border} and {\it noise points}  will read as:
 
 \begin{itemize}[leftmargin=0.5cm]
	\item \cbf{core point}: if  $M_\varepsilon(k,l)\geq K$ .
 	\item \cbf{border point}:  if  $M_\varepsilon(k,l)< K$, but
 	at least one core point belongs to  $B_{\varepsilon}(k,l)$. 
 	\item \cbf{noise point}: if both  the conditions above are not meet.
 \end{itemize}

The choice to use as DBSCAN scanning brush a box rather than a circle, speeds-up the computational time, indeed we don't need to evaluate the CCD  pixels  distances from  $p_{k,l}$ , but just to  slice the sub-matrix corresponding to $B_{\varepsilon}(k,l)$. We use values of $\varepsilon$ of a few pixels, typically 1. 
The remaining part of the algorithm, concerning the recursive build-up of the clusters, follows the original implementation.

\begin{figure*}
	\includegraphics[width=15cm] {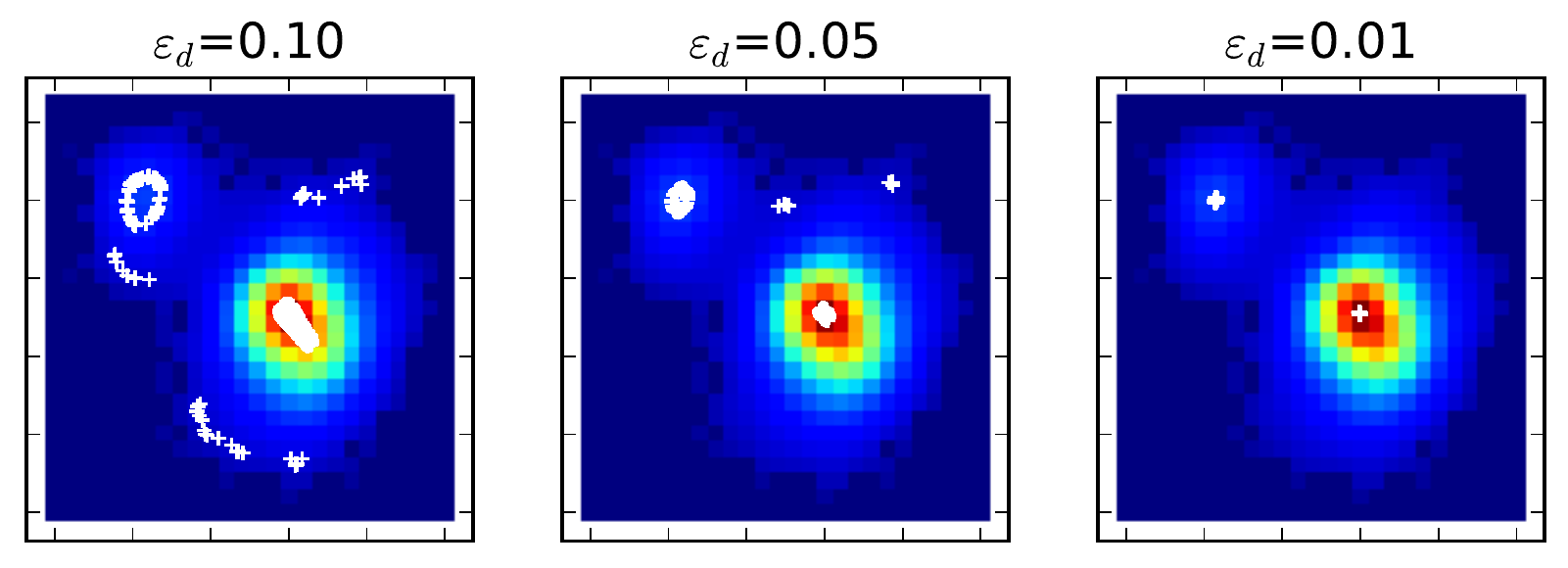}
	\caption{Distribution of the density attractors (white crosses) for the cluster with ID=0 in central panel of Fig. \ref{fig:fig_deblending} for different values 
		of $\varepsilon_d$  (see Sec. \ref{sec:deblending} ).  Lower  values of $\varepsilon_d$ result in a  tighter clustering of 		the attractors toward the local maxima of the image.  The input  image is a {\it gri} summed bands
			image  cutout centered on  the object with DR8OBJID=1237663548511748377 from the Galaxy Zoo 2 Main sample with spectroscopic redshifts. }
	\label{fig:fig_attractors_evolution}
\end{figure*}

 \begin{figure*}
 	\includegraphics[width=15cm]{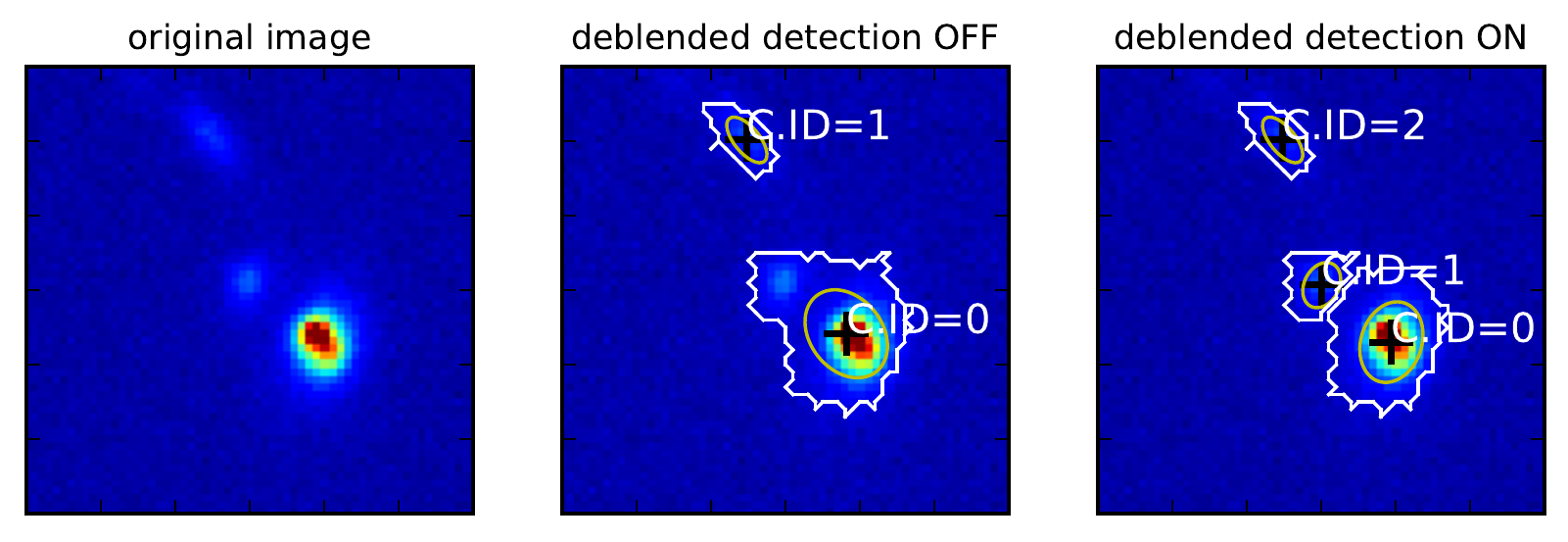}
 	\caption{Application of the DENCLUE algorithm to deblend 
 		two confused sources (see Sec. \ref{sec:deblending}): original image (left), DBSCAN detection result (center), 
 		and result of the detection after DENCLUE-based deblending (right). Source and image provenance are the same as in Fig. 3.}
 	\label{fig:fig_deblending}
 \end{figure*}

In order  to  speed-up further the computational time, we  have implemented in our version of the algorithm the possibility to remove from the  iteration all the CCD pixels with a flux below a given background threshold. The background threshold is evaluated using the following method:
\begin{itemize}[leftmargin=0.5cm]
	\item We split the CCD matrix  in N sub-matrices  (typically N=10).
	\item We select the sub-matrix with the  lowest integrated flux,  and we estimate the mode  of the flux distribution $m_{bkg}$, and it's standard deviation $\sigma_{bkg}$. 
	\item We compute a range of skewness values for distributions obtained by excluding flux values outside $n \sigma_{bkg}$ from the mode, where $n$ range from 0.5 to 2.0 with 0.1 step. We retain the $n$ value which leads to the lowest skewness and call it n*.
	\item   $m_{bkg}$ and $\sigma_{bkg}$ are updated for the new flux distribution sigma-clipped for $n^*$
	\item We set the  the pre-filtering background value  $bkg_{th}$ at  the level of $bkg_{pre-th}=N_{bkg}\sigma_{bkg}+m_{bkg}$ above the mode.
	The value of  $N_{bkg}$ is chosen in the range of $\simeq$ [2.5,3.5], in order to remove the bulk 
	of the background points, but leaving at  the same time enough statistic for the noise 
	determination embedded in the DBSCAN.
	\item We apply a boolean mask to all the CCD pixels with a flux below $bkg_{pre-th}$.
\end{itemize}
A schematic view of the image segmentation process in shown in the top box of Fig. \ref{fig:detection_flwchart}, and an example of the pre-filtering procedure is shown in the left and central panels of  Fig. \ref{fig:fig_dbs_detection}. 
The left pane of Fig. \ref{fig:fig_dbs_detection}.  shows  the original image, and the white dots in the central panel of the same figure show the pixel selected  after the background-based pre-filtering.

The last step, to apply the DBSCAN algorithm, is the setup of the parameters $K$, i.e. the DBSCAN threshold, and 
$\varepsilon$ i.e. the half width of the DBSCAN scanning box..
The parameter $K$ is the one tuning the DBSCAN internal noise determination, and we use the background pre-filtering to set the value of $K$ using the following method:

\begin{itemize}[leftmargin=0.5cm]
	\item We set $K=K_{th}bkg_{pre-th} $ where the parameter $K_{th}$ is typically in the range 
	of [1.0,2.0].
	\item To avoid that  the value of $K$ depends on $\varepsilon$,  
	$M_\varepsilon(p_{k,l})$ is averaged over the number of pixels in the DBSCAN scanning box. 
	In this way the value of $K$ represents a {\it per-pixel} threshold.
\end{itemize}
The second parameter, $\varepsilon$, is the one which tunes the size of the DBSCAN scanning 
box, hence, low values of $\varepsilon$ will allow to follow accurately also the contour 
of small objects. For this reason, throughout the present work, we have used $\varepsilon=1.0$ pixels,  meaning that the scanning box will have a size of 9 pixels. 
For each source cluster we evaluate the following relevant parameters:
\begin{itemize}[leftmargin=0.5cm]
	\item[-] $(x_c,y_c)$ the centroid coordinates.
	\item[-] The cluster containment ellipsoid defined by  the major and minor semi axis  $\sigma_{x},\sigma_{y}$, and by the inclination angle  $\alpha_{\rm PCA}$, measured counterclockwise angle w.r.t.  the $x$ axis.
	All these parameters are evaluated  by applying the principal component  analysis method (PCA) \citep{Jolliffe1986},
	to the covariance matrix of the arrays of the cluster point position $\mathbf{x}$, $\mathbf{y}$, weighted on the cluster pixel flux. 
	This method	uses the eigenvalue decomposition of the covariance matrix of	the two position arrays $\mathbf{x}$ and  $\mathbf{y}$. By definition, the square root of the first eigenvalue will correspond to $\sigma_{x}$, and the second to $\sigma_{y}$.
	\item[-] $cnt$  the set of the  coordinates of the cluster edges pixels 
	\item[-] $r_{pca}=\sqrt{\sigma_{x}^2 + \sigma_{y}^2}$ 
	\item[-] $r_{max}$ , i.e. the distance,from the cluster centroid, of   the most distant cluster pixel. 

\end{itemize}

The right panel In Fig. \ref{fig:fig_dbs_detection}  shows the final result for an image from our data set, 
with  $N_{bkg}=3.0$, $K_{th}=1.5$ and $\varepsilon=1.0$. The white lines represent 
the edges pixels of the clusters (sources), and  the yellow ellipses represent the cluster containment ellipsoid, defined by    $\sigma_{x},\sigma_{y}$, and $\alpha_{\rm PCA}$.
The black crosses represent the clusters centroids.

\subsection{Source deblending: DENCLUE}
\label{sec:deblending}
When sources are very close, and/or when we need to use a low value of $K_{th}$ in order to recover  faint 
structures, as in the case of detection of spiral arms, it might happen that the DBSCAN algorithm is 
not able to  separate them (see Fig. \ref{fig:fig_deblending},  central panel).

To deblend 	 two (or more) `confused' sources we have implemented a deblending method based on  the  DENCLUE algorithm  \citep{Hinneburg:1998wo,Hinneburg:2007tq,Zaki:2014}. 
The original implementation of the DENCLUE algorithm is based on the kernel density estimation to find local maxima of dense region of points.
In the case of digital images it is not possible to apply straightforwardly the equations reported in the original algorithm implementation, indeed the pixels coordinates have a uniform spatial distribution. To overcome this limitation we have modified the DENCLUE algorithm substituting the kernel density estimation
with a convolution of the image with a given kernel. Let $p_j$ be the $j_{th}$ pixel with coordinates
$\mathbf{q_j}$, the kernel function   $G$ is a non-negative and symmetric function,  
centered at $\mathbf{q_j}$ that represents  the influence of the pixel with $p_i$ on $p_j$. The  convolved image at $p_j$ is estimated by the function $f$ as:
\begin{equation}
f(p_j) \propto \sum_{i=1}^{n}  G\big(\frac{\mathbf{q}_j-\mathbf{q}_i}{h} \big) I(p_i)
\end{equation}
where  $n$ is the number of pixels in the domain of the function $G$, $h$ is the bandwidth of the kernel, and $I(p_i)$ is the 
image flux at the pixels $p_i$ with coordinates $\mathbf{q}_i$. 
For example, in the case of a two dimensional  Gaussian kernel, the function $G$ will read:
\begin{equation}
G(\mathbf{q}) \propto \exp(\mathbf{-z z^T}),	
\label{eq:gauss_kernel}
\end{equation}
where:
\begin{align*}
\mathbf{z}=\frac{\mathbf{q-q}_i}{h} 
\end{align*} 
and the bandwidth of the kernel $h$, acts as the standard deviation of the distribution.
The DENCLUE algorithm is designed to find for each
point $p_j$ the corresponding \cbf{density attractor} point i.e a local maximum of $f$.
To find the attractors,  rather than using a computationally expensive gradient  ascent approach, we use the fast hill climbing technique presented in  \cite{Hinneburg:2007tq} and \cite{Zaki:2014}, that is an iterative  update rule with the formula:
\begin{equation}
\mathbf{q}_{t+1}=\frac{ \sum_{i=1}^{n} G\big(\frac{\mathbf{q}_t-\mathbf{q}_i}{h}\big)\mathbf{q}_i I(p_i)}
{ \sum_{i=1}^{n} G\big(\frac{\mathbf{q}_t-\mathbf{q}_i}{h}\big)I(p_i)}
\end{equation}
where the $t$ is the current iteration, $t+1$ the updated value, and $\mathbf{q}_{t=0}\equiv\mathbf{ q}_j$.

The fast hill climbing starts at each point with coordinate vector  $\mathbf{q}_j$, 
and iterates until $\|\mathbf{q}_t -\mathbf{q}_{t+1} \| \leq \varepsilon_d$. 
The coordinate  vector $\mathbf{q}_{t+1}$ identifies  the position of the \textit{density attractor}
$p_j^*$ for the   point $p_j$. 

\begin{figure*}
	\centering
	\begin{tabular}{lll} 
		\includegraphics[height=8.2cm]{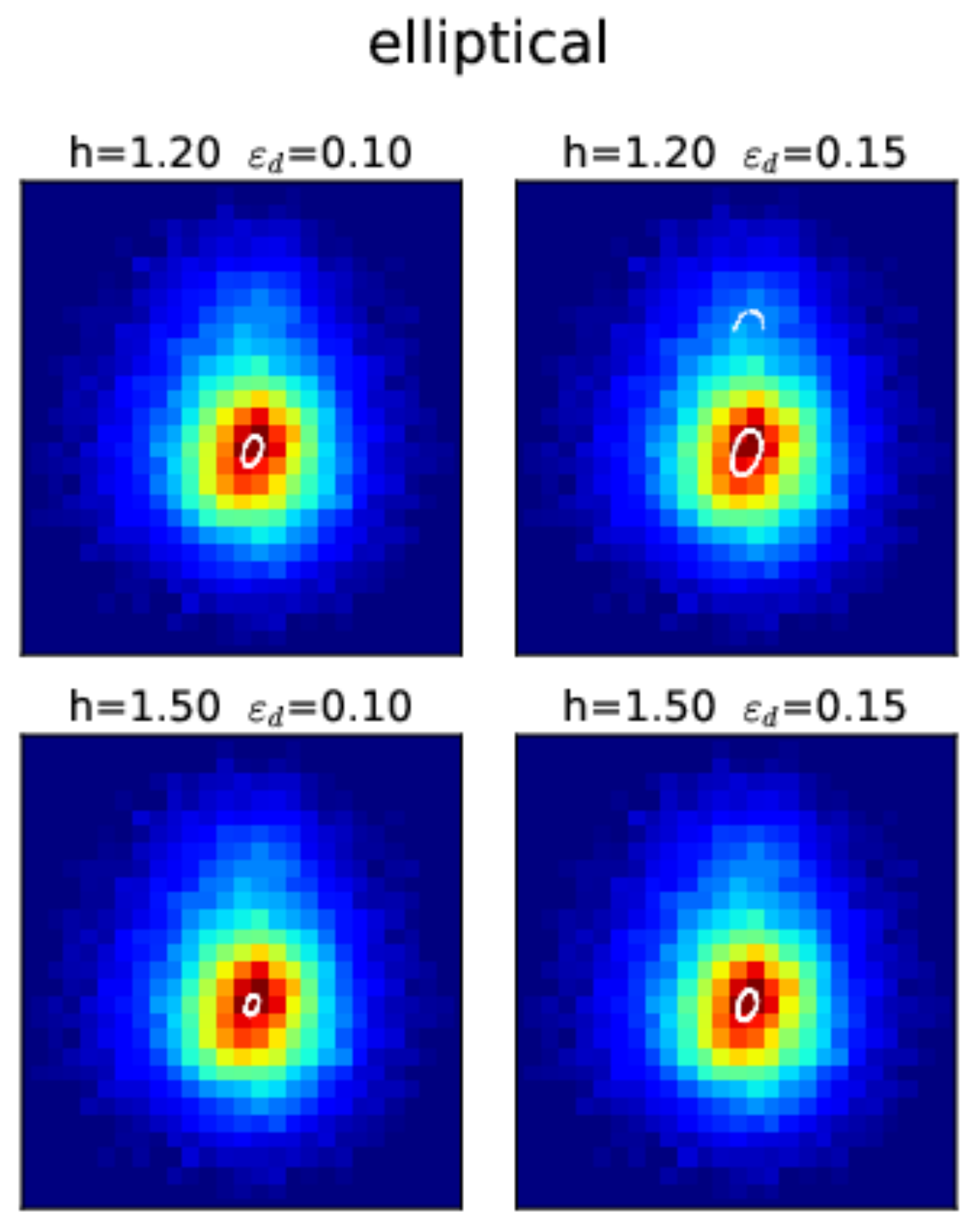}& &
		
		\includegraphics[height=8.2cm]{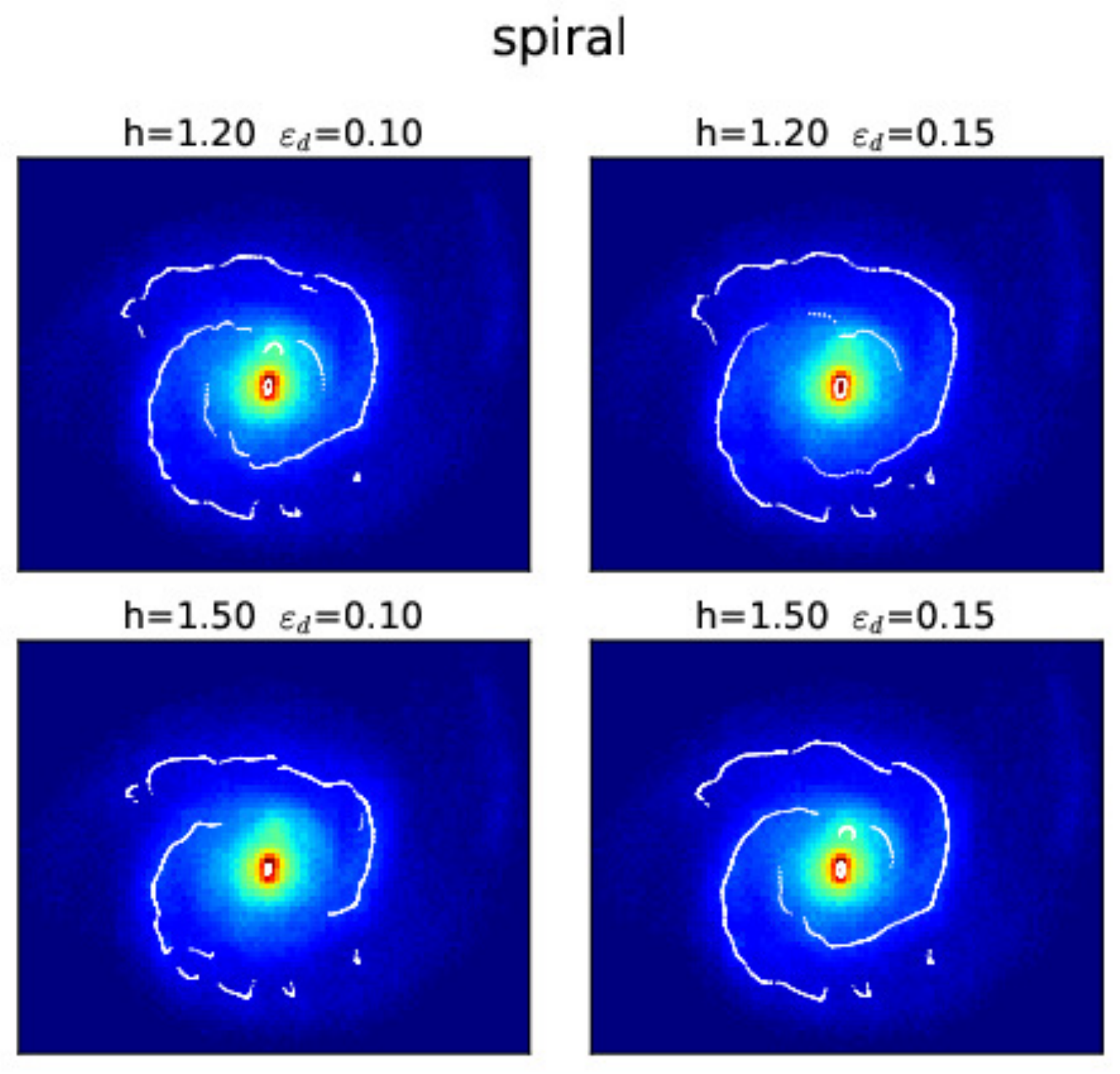}\\
		
	\end{tabular}
	\caption{Application of the DENCLUE algorithm to spiral arms tracking (see Sec. \ref{sec:denclue_spir_arms}). White dots represent density attractors.
		\textit{Left panels}: density attractor for an elliptical galaxy, for different values of $h$, and $\varepsilon_d$, reported in the figures. \textit{Right panels:} same as in left panels, for a spiral galaxy.		
		The images correspond to a {\it gri} summed bands cutout centered on the object with ID DR8OBJID=1237657233308188800 from the Galaxy Zoo 2 SDSS Stripe 82 sample (left panels), and
			DR8OBJID=1237659756599509101 (right panels).
	}
	\label{fig:fig_denclue_spiral_arms}
\end{figure*}

Once that all the attractors have been evaluated, then  they  are  clustered using the DBSCAN algorithm  to eventually deblend the confused sources, in a way that can be summarized by the following steps:
\begin{itemize}[leftmargin=0.5cm]
	\item each source cluster $S$, detected by the DBSCAN, is defined by a set of 
	points $\{p_j \in S\}$, corresponding the pixels in the source with coordinates $\mathbf{ q}_j$,
	and flux value $I(p_j)$
	\item for each point $p_j \in S$  we compute the  \textit{density attractor}. 
	It means that for each pixel $\forall p_m \in S, \exists p_m^*$, i.e. the two sets $\{p_m \}$ and $\{p_m^*\}$, map bijectively $S$ to $S^*$.
	\item all the attractors points  $S^*$ are clustered using the DBSCAN, in clusters
	\textit{density attractors} producing a list of clusters of attractors  $\{CA_1,...,CA_n\}$.  
	\item The set of pixels  $p_m$ whose  \textit{density attractors} belong to the same cluster of 
	attractors $CA_n$, i.e. $\{p_m:p^*_m \in CA_n \}$, defines a  new sub-source $s_n$ 	(sub-cluster)
	\item Each sub-source $s_n$ can eventually be validated or discarded according to some criteria:
	\begin{itemize}
		\item minimum pixels  number
		\item maximum pixels number
		\item ratio of the sub-cluster flux compared to the parent source flux 
	\end{itemize}
\end{itemize}

A schematic view of the DENCLUE-based deblending  process is shown in the bottom box of Fig. \ref{fig:detection_flwchart}

In the following we will use  a Gaussian kernel function for the DENCLUE-based source deblending. 
We note that both  $h$ and   $\varepsilon_d$ have a significant impact on the final result 
of the deblending.
The kernel width $h$ is relevant to the `scale' of the deblended sub-clusters, indeed large values of $h$ will smooth high frequency signals, on the  contrary small values will preserve small scale features. 
We have found that a value of  kernel width $h$ of the order of   $\simeq 0.1\times r_{max}$, 
provides good results in deblending sources, avoiding to fragment objects with complex morphology  (such as spiral galaxies), and in separating close sources, even with a large difference in the integrated  flux. 
The parameter  $\varepsilon_d$ is responsible for the convergence of the fast hill climbing algorithm, hence 
for the determination of the position of the final attractor. A large value of  $\varepsilon_d$, will allow to 
find local maxima related to noisy pixels, or morphological  features of the source, a small value, on the contrary, will lead to 
track more significant maxima related to the core of the galaxy. We can see this clearly in Fig. \ref{fig:fig_attractors_evolution}, where  we show the \textit{density attractors} (white crosses) for different values 
of  $\varepsilon_d$.  We note that,  as $\varepsilon_d$ is decreasing, all the 
attractors gets more and more tightly clustered around the two local maxima, corresponding to the two source cores.

Finally, in Fig. \ref{fig:fig_deblending} we show how the deblending algorithm works. The left panel 
shows an image with three   sources, two of which separated by a few pixels. The central panel shows the 
source detection with the DBSCAN threshold set to  $K_{th}=1.5$, which finds only two  sources.
The  right panel shows the image after the application of the DENCLUE-based deblending method with a
 Gaussian kernel function, with $h=$0.05$~r_{max}$.

\section[]{application of DENCLUE to track spiral arms }
\label{sec:denclue_spir_arms}
Since the DENCLUE algorithm is able to track flux maxima in the 2D   images, we decide
to test whether this capability is able to track spiral arms too.
As a first step we need to distinguish between cluster of \textit{density attractors} related to the  core of the galaxy, 
i.e. \cbf{`core' density attractors} clusters, and \cbf{ `non-core'  density attractors} clusters, that could be related to spiral arms patterns. 
We define  the  \textit{`core' density attractors } cluster as the cluster of attractors with the smallest distance form the galaxy centroid $(x_c,y_c)$, and fully contained within  the galaxy 
effective radius $r_{eff}$. 
	
The second step  is to find the optimal configuration of the DENCLUE parameters for tracking spiral arms, i.e. we face a situation that is opposite to  that of 
source debelnding. Indeed, in this case we are interested in finding attractors not only related to 
the core of the source, but also to fainter morphological features.
As anticipated in the previous section, the bandwidth $h$, and the fast hill 
climbing threshold $\varepsilon_d$ play a relevant role in the  extraction 
of the \textit{density attractors}. Of course, $h$ has to be comparable with the 
scale of the feature that we want to extract, while	 $\varepsilon_d$ will tune 
the impact of the level of noise on the detection of the attractors. 
We have found that  a Gaussian kernel with a bandwidth $h$ in the range [1.0,2.0]
is able to track well spiral arms features, for the largest fraction of source sizes in our dataset, and  that the  optimal choice of  $\varepsilon_d$ is in the range [0.1,0,2].

Even though we have identified an optimal range for both the two DENCLUE parameters, still
it can happen that a given combination of $\varepsilon_d$ and $h$ is not able to 
find a cluster of attractors that  meets the requirements to be a `core' \textit{density attractors} cluster as in the  case of very noisy images.
In order to mitigate such a possible bias we have implemented and automated 
iterative procedure to set the values of the two parameters of interest,  $\varepsilon_d$ and  $h$.
The iteration is performed decreasing  $\varepsilon_d$ of 5\%, and increasing $h$ of 5\%, until the `core' cluster of  \textit{density attractors}  has been found, 	 or a maximum of $maxtrials = 10$ iterations is reached.

An example of this application is given in  Fig. \ref{fig:fig_denclue_spiral_arms} where we show 
 the typical  case for an elliptical  (left panels) and a spiral galaxy (right panels), for $h=1.2,1.5$ , and  $\varepsilon_d=0.10,0.15$.
It is clear that in the case of the elliptical galaxies  `non-core'
\textit{density attractors} are absent or  rare, on the contrary, 
in the case of a spiral objects, we note a significant number of `non-core'  \textit{density attractors}, 
following quite well the local maxima of the image related to the spiral arms.

\begin{figure}
	\centering 
	\includegraphics[width=8.5cm]{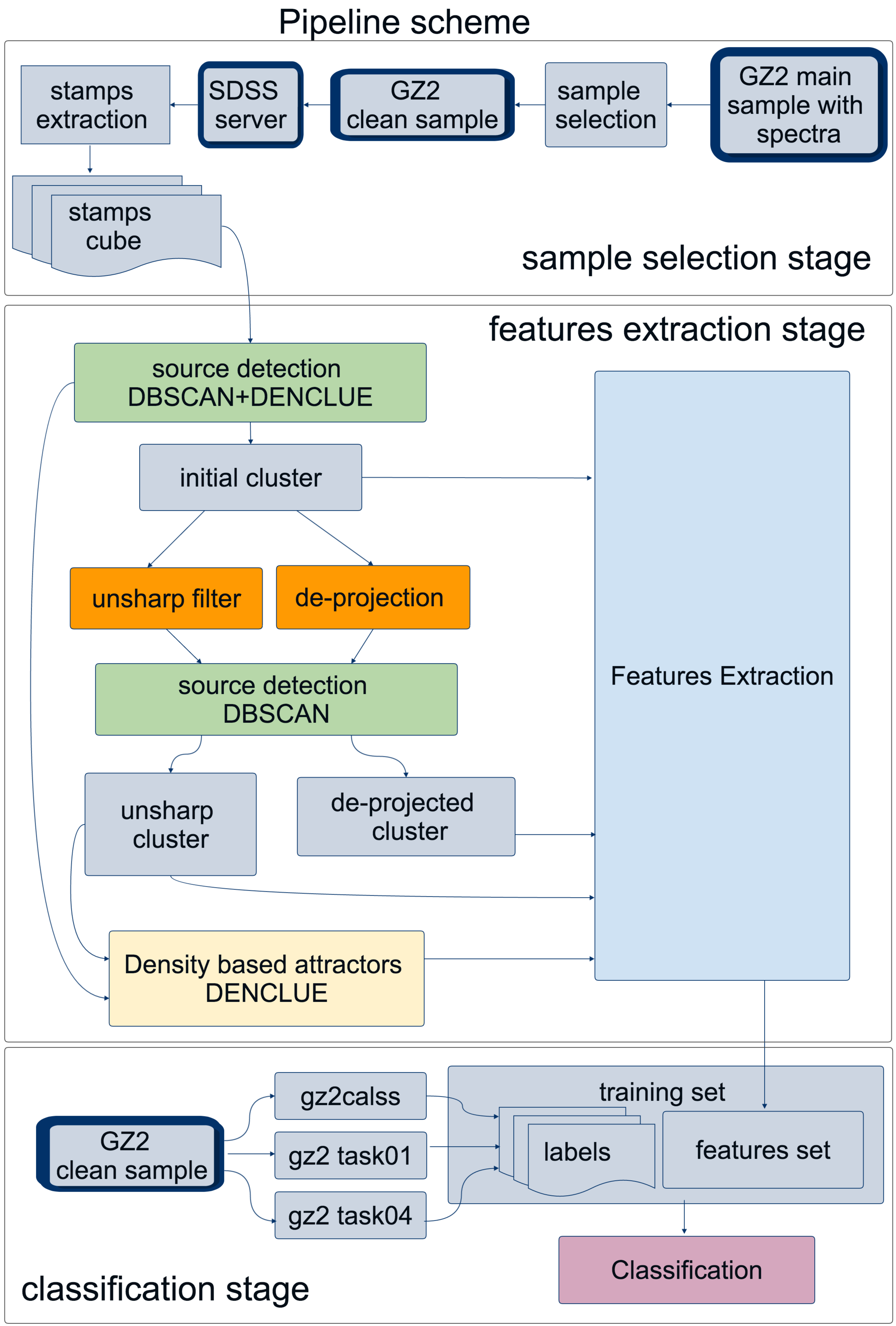}
	\caption{Flow chart diagram showing the structure of the data processing pipeline}
	\label{fig:fig_pipeline}
\end{figure}

\section{The  ASTErIsM pipeline for the automatic galaxy shape identification}
\label{sec:pipeline_general}
The  \texttt{ASTErIsM} pipeline for the automatic galaxy shape identification consists of three different stages:
\begin{enumerate}[leftmargin=0.5cm]
	\item Sample selection.
	\item Features extraction.
	\item Classification.
\end{enumerate}
a schematic view  is shown in Fig. \ref{fig:fig_pipeline}, and technical details about the code implementation are reported  in App. \ref{App:Tech_details}.
In the following of the paper we will describe  each stage of the pipeline,  the extracted 
products, and their impact of the morphological classification.

  \begin{table}
  	
  	\centering
  	\caption{GZ2 decision tree tasks used in our sample selection}
  	\label{tab:gz2_tasks}
  	\begin{tabular}{|c|C{4cm}|l|}
  		\hline
  		Task & question & Answer  \\
  		\hline
  		\multirow{ 3}{*}{01}   & \textit{Is the galaxy simply smooth and rounded, with no sign of
  			a disk ?} & A1.1 smooth \\
  		& & A1.2 feature or disk\\
  		& & A1.3 star or artifact \\
  		\hline
  		\multirow{2}{*}{02}   &  \textit{Could this be a disk viewed edge-on  ?} & A2.1 yes \\
  		& & A2.2 no\\
  		\hline
  		\multirow{2}{*}{04}   &  \textit{Is there any sign of a spiral arm pattern  ?} & A4.1 yes \\
  		& & A4.2 no\\
  		\hline
  	\end{tabular}
  \end{table}

\subsection{Sample selection}
\label{sec:sample selection}
The galaxy images and the morphological classification have been 
taken from the Galaxy Zoo 2 \citep{Willett:2013ea} (GZ2) SDSS Main sample  with spectroscopic redshifts. \footnote{http://zooniverse-data.s3.amazonaws.com/galaxy-zoo-2/zoo2MainSpecz.fits.gz} 
 We have identified three relevant questions (or tasks) from the GZ2 decision tree \citep{Willett:2013ea}, reported in Tab. \ref{tab:gz2_tasks}, that are useful
 to identify a sub sample of well-identified spiral arms and elliptical objects.
 For each answer we have chosen, as decision variable, the \textit{debiased} fraction of votes, that measures the agreement between a single user  vote, and the best answer to that question   \citep{Willett:2013ea}.
 The following selection criteria have been used to  select from GZ2  a sample of clean elliptical and clean spiral  objects (see Tab. \ref{tab:gz2_tasks}) :

 \begin{itemize}[leftmargin=0.5cm]
 	\item Elliptical galaxies cut : \texttt{ t01, A1.1}$>0.9$ 
 	
 	\item Spiral galaxies cut:  \texttt{ t01, A1.2}$>0.9$  \&\&  \texttt{ t02, A2.2}$>0.9$ \&\& \texttt{t04, A4.1 }$>0.9$
 \end{itemize}
producing a final  \textit{clean} sample of 24635 objects.
Each object in sample  has been labeled according to the class extracted from the column \texttt{gz2class}  in the GZ2 catalog. All objects with the \texttt{gz2class}  string starting with `E'  have been labeled as elliptical,  and all the objects with string starting with `S' have been labeled as spiral.  In total we have 7186 elliptical objects and 17449 spiral objects.
  
 For each object we have downloaded the \textit{g,r} and \textit{i} bands fits image from the SDSS,   and a stamp of 100 by 100 pixels has been extracted, summing the three bands images, and centered on the source  coordinates reported in the GZ2 catalog.
 We have decided to work with FITS image format, because compressed format such as JPEG, 
 present interpolation features that might lead to spurious  \textit{density attractors}.

\begin{table*}
	
	\centering
	\caption{Detailed description of the features. The first column 
		reports the kind of cluster used to extract the image, the second the category of the features (clustering/morphology). 
		The third column reports the name of the features group, and in round brackets is reported 
		the section of the paper where the features are presented.
		The fourth columns reports the root name of the feature. The fifth column reports a flag, added to the 
		name, with \texttt{\_ic} corresponding to the \textit{initial cluster}, \texttt{\_depr} corresponding
		to the \textit{deprojected} cluster, and \texttt{\_unsh} corresponding to the \textit{unsharp} cluster.
		The last column reports the number of features $N_f$ for each group.}
	\label{tab:features}
	\begin{tabular}{clllll}
		\hline
		Cluster & Category & Group (Sec.)  & var. root name& flag & $N_f$  \\
		\hline 
		\multirow{ 8}{*}{initial}&\textit{clustering}    &		  Geometrical features (\ref{sec:geom_features}) & \texttt{geom}&	\multirow{ 8}{*}{\texttt{ic}}  & 7 \\
		
		&\textit{clustering}   &		  Hu moments   of cl. contour  (\ref{sec:Hu_moments})   &\texttt{cnt\_log\_Hu}   &     &  7\\
		&\textit{clustering}	 &  Hu moments   of cl. image   (\ref{sec:Hu_moments})    			 &\texttt{cl\_img\_log\_Hu}  &   &  7\\
		&\textit{clustering} &		  Hu moments  of  \textit{density attractors}  (\ref{sec:attractors_features})      & \texttt{attr\_log\_Hu}  & &  7\\
		&\textit{clustering} &		  Hu moments  of  \textit{density attractors}, polar coord.  (\ref{sec:attractors_features})     & \texttt{attr\_polar\_log\_Hu}  & &  7\\
		&\textit{morphology} & Morphology of cl. image     (\ref{sec:Morph_features})               &\texttt{morph}   &	& 10\\
		
		\hline
		\multirow{ 2}{*}{unsharp}&\textit{clustering}    &		  Geometrical features (\ref{sec:geom_features}) & \texttt{geom}&	\multirow{ 2}{*}{\texttt{unsh}}  & 7 \\
		
		&\textit{morphology} & Morphology of cl. image     (\ref{sec:Morph_features})               &\texttt{morph}   &&	10\\
		
		\hline
		\multirow{ 4}{*}{deprojected}&\textit{clustering}    &		  Geometrical features (\ref{sec:geom_features}) & \texttt{geom} &	\multirow{ 4}{*}{\texttt{depr}}  &7 \\
		
		&\textit{clustering}   &		  Hu moments   of cl. contour  (\ref{sec:Hu_moments})   &\texttt{cnt\_log\_Hu}        & &7\\
		&\textit{clustering}	 &  Hu moments   of cl. image   (\ref{sec:Hu_moments})    			 &\texttt{cl\_img\_log\_Hu}     & & 7\\
		
		&\textit{morphology} & Morphology of cl. image     (\ref{sec:Morph_features})               &\texttt{morph}   &	&10\\
		
		\hline	
		initial+ &\textit{clustering}    &	 Radial distribution of  non-core \textit{density attractors} (\ref{sec:attractors_features}) & \texttt{r\_distr\_attr}   & 	\multirow{ 2}{*}{\texttt{ic}} & 7	 \\
		unsharp &\textit{clustering}    &	 Angular distribution of  non-core \textit{density attractors} (\ref{sec:attractors_features}) & \texttt{theta\_distr\_attr} &   & 5	 \\				
		\hline	
	\end{tabular}
\end{table*}

\subsection[]{Features Extraction}
\label{sec:features_pipeline}

The features extraction stage consists of 4 main steps, and for each step a set of features is extracted, a schematic view of the stage is shown in the features extraction box of Fig. \ref{fig:fig_pipeline}.
\begin{itemize}[leftmargin=0.5cm]

\item \cbf{initial clusters}: The first step  is used to detect the central source in the stamp, and to de-blend from possible nearby   
contaminating sources.
If the source cluster has  $r_{max}$  larger than 35 pixels, then the image stamp is rescaled in order to produce  a source cluster with $r_{max}\leq35$.
We have checked that this rescaling has no impact on the quality of the  extracted features, and has the only motivation to reduce the average computational time to $\simeq3.5$ seconds per stamp.
 
The source cluster produced at this stage is labeled as\cbf{initial cluster}, and the corresponding features   are  flagged with  \texttt{\_ic} string. We extract from this cluster a `cluster image' by building  an image 
with the same size  of the original stamp, with null pixels, and setting the pixels belonging to the source cluster,  to their actual flux value. This `cluster image' is used as input for the extraction of the `de-projected' cluster and of the `unsharp' cluster', and to extract image-related features.

\item \cbf{de-projected clusters}: To extract the  `de-projected' cluster we de-project the \textit{initial cluster} `cluster image' by rotating and applying an affine  transformation in order to map the galaxy shape to a shape that is as close as  possible to circular.  The re-pixelization of the image is based  on 
 third order spline interpolation, provided by the \texttt{ndimage} module from  the \texttt{scipy} package \citep{scipy}. The detection process is again performed on this final image, and the `de-projected'  cluster with corresponding `cluster image' are produced. The features are extracted and flagged with the string \texttt{\_depr}

\item \cbf{unsharp clusters}:  To extract the  `unsharp' cluster we apply an unsharp filter to the  \textit{initial cluster} `cluster image'.  This kind of filter is useful  to enhance edges and high frequency features of the image. It is based on the subtraction  between of a smoothed version  of the original image, from the original one, producing 
and unsharp filtered image:
\begin{equation}
	I^*(x,y)=I(x,y) - u\cdot I_{smooth}(x,y)
\end{equation}
where the image is smoothed by  a Gaussian filter with bandwidth $\sigma_s$,
and the subtraction level is tuned by the parameter $u$. 
The value of $\sigma_s$ is chosen to be proportional to the Petrosian radius 
of the source (see \ref{sec:Morph_features} for a definition of the Petrosian radius).
The detection process is again performed on the unsharp image,  and the `unsharp'  cluster with the corresponding `cluster image' are produced. The features are extracted and flagged with the string \texttt{\_unsh}.

\item \cbf{density attractors} As final steps we extract the  \textit{density attractors} from the `initial cluster', using 
the pixels that are common both to the `initial' and to the `unsharp' cluster, i.e. the convolution equation for the DENCLUE algorithm will read:
\begin{equation}
	f(p_j) \propto \sum_{i=1}^{n}  G\big(\frac{\mathbf{q}_j-\mathbf{q}_i}{h} \big) I(\mathbf{q}_i)
\end{equation}
where $j \in$ `initial cluster', and $i \in$ `unsharp cluster'. This choice allows to avoid that some
attractors will match the edges of the source clusters, lowering the possibility to detect 
attractors not related to spiral arms structures.
  \end{itemize}

We will distinguish the extracted features in two categories:
 
\begin{itemize}[leftmargin=0.5cm]
\item \cbf{clustering-related}: These features can be  evaluated only if  a cluster 
is extracted, and are evaluated from the  information stored in the 
source cluster, or  from the extracted cluster image.

\item \cbf{morphological}:  These features have already been used in the literature (see Sec. 
\ref{sec:Morph_features}), and could be evaluated without the need 
to extract the source cluster.
\end{itemize}

All the features, are summarized in Tab \ref{tab:features}. The feature name is built according to the following scheme:  \texttt{root\_name+specific\_name+flag}. In the following,  if not specified otherwise, we will refer to the specific name.

\subsubsection{Geometrical Features}
\label{sec:geom_features}
 Since the DBSCAN cluster shape is arbitrary, i.e. there is no constraint coming from convolution 	with a predefined shape, the extracted cluster preserve as much as possible 	the actual shape of the source. For this reason, we have identified a set of features that  maximize the cluster geometrical information, measured from the position of the source cluster points, and of the contour points:

\begin{itemize}[leftmargin=0.5cm]
	\item \texttt{geom\_pix\_size} -  the number of pixels  of the cluster.  
	\item \texttt{geom\_r\_max} -  the $r_{max}$ of the cluster.
	\item \texttt{geom\_ecc} - the eccentricity of the cluster containment ellipsoid (see 
	Sec.\ref{sec:source detection}).
	\item \texttt{geom\_comp} - the geometrical compactness,  defined as $P^2/A$, where $P$ is the contour 
	perimeter (i.e. the number of pixels flagged as contour), and $A$ is the 
	total number of pixels of the cluster.
	\item \texttt{geom\_ar} - the aspect 
	ratio of the minimal rectangular box enclosing the source cluster.
	\item \texttt{geom\_contour\_ratio} -  the contour ratio defined as  $P_{box}/P$, where $P_{box}$ is the contour 
	perimeter of the minimal  rectangular box enclosing the cluster.
	\item \texttt{geom\_area\_ratio} - the area ratio, defined as $A_{box}/A$, where $A_{box}$ is the area of 	the minimal rectangular box enclosing the cluster.
\end{itemize}

\subsubsection{Invariant Hu moments }
\label{sec:Hu_moments}

 We can increase the amount of information provided by the DBSCAN clusters,  by evaluating the  Hu moments \citep{Hu:1962}, which 
are  a set of 7 geometrical moments that are invariant under scaling, rotation,  
and translation. The complete  definition is reported  in Appendix \ref{App:Hu_moments}.
The following features are extracted:
\begin{itemize}[leftmargin=0.5cm]
\item \texttt{cnt\_log\_Hu\_[0-6]} - the logarithm of Hu moments of the cluster contour.
\item \texttt{cl\_img\_log\_Hu\_[0-6]} - the logarithm of Hu moments of the cluster image.
\end{itemize}

\begin{figure*}
	\centering
	\begin{tabular}{ll} 
		\includegraphics[width=8.3cm]{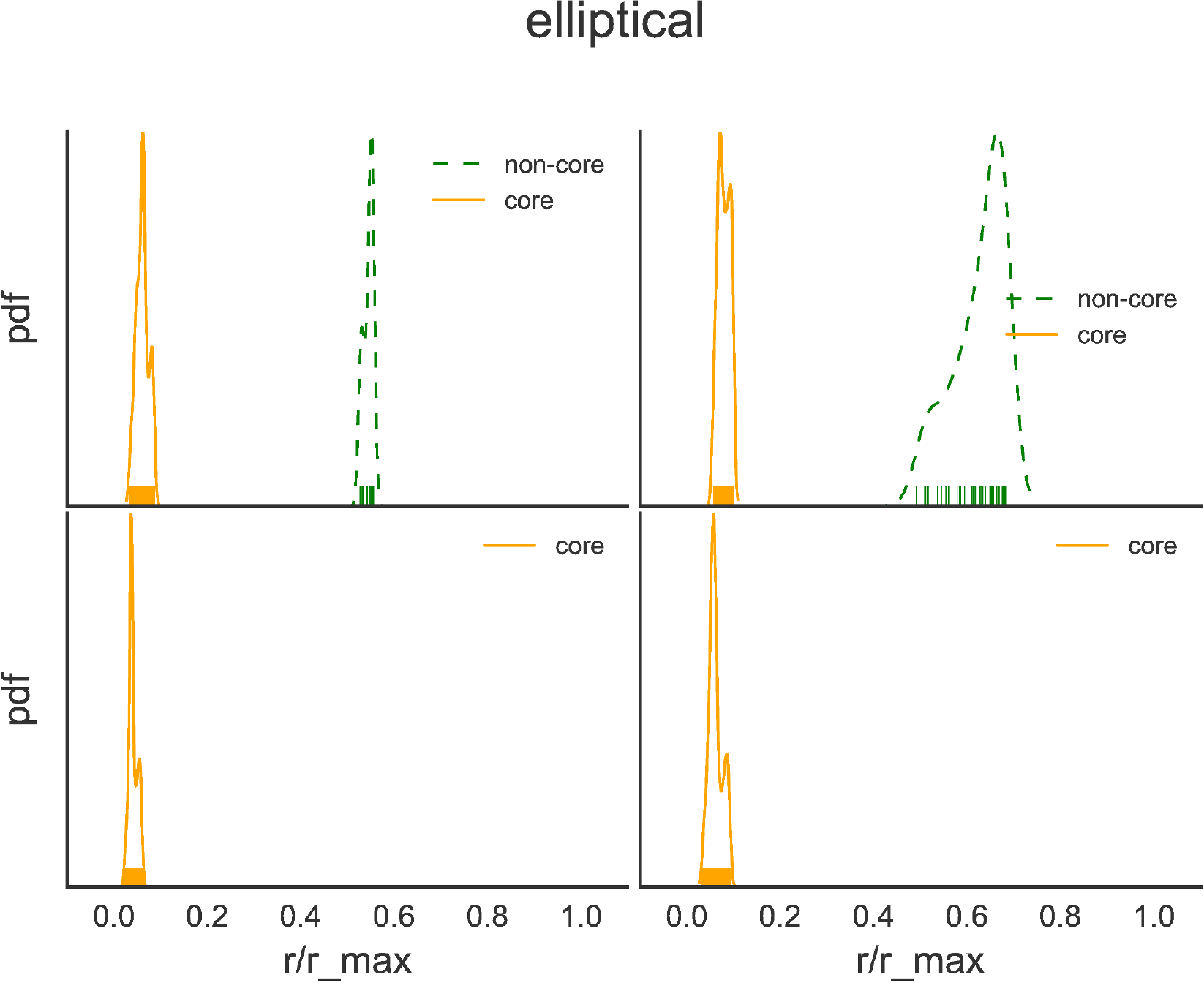}&
		\includegraphics[width=8.3cm]{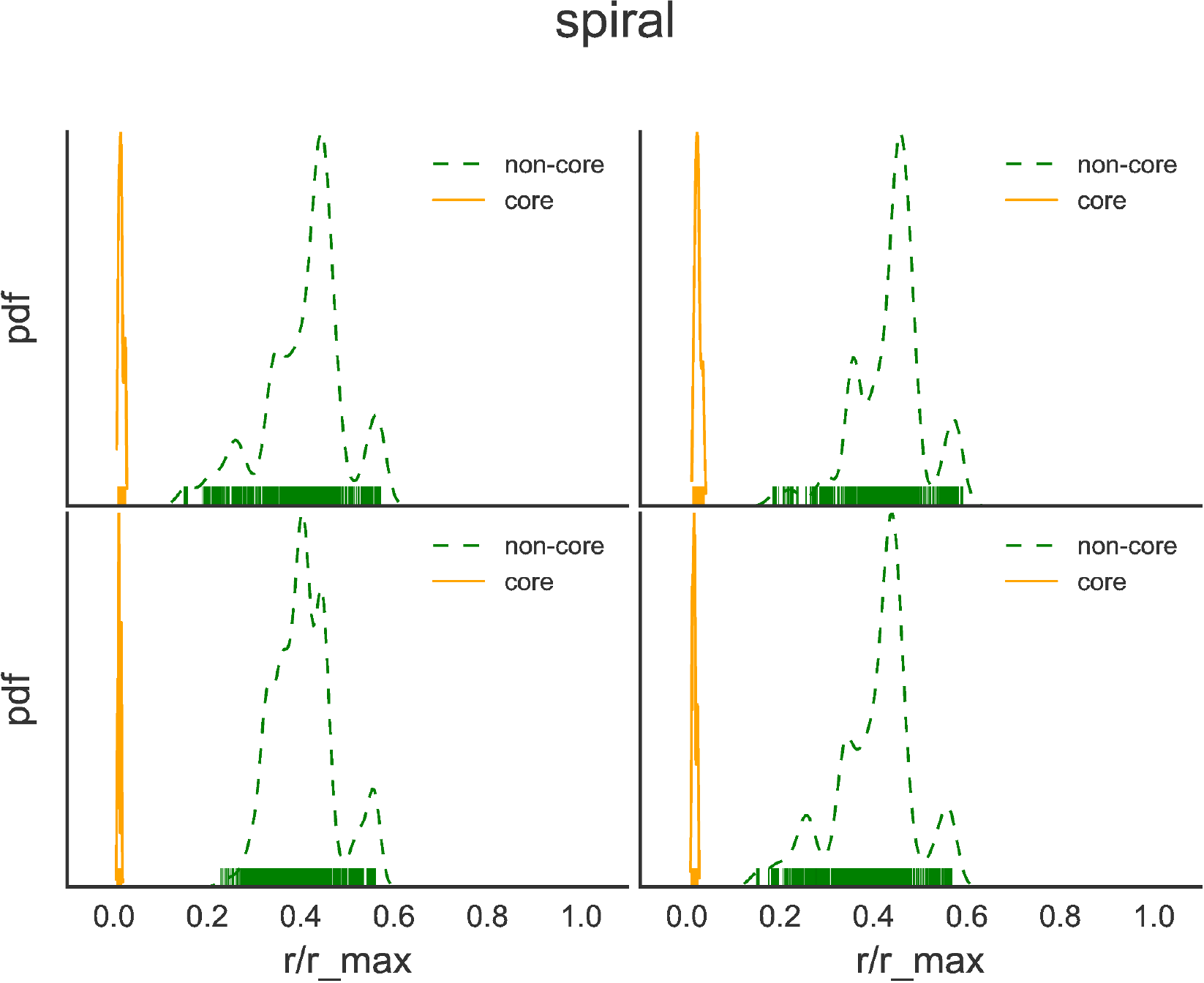}\\
		
	\end{tabular}
	\caption{The radial distribution of \textit{the density attractors} for the detections 
		of the two galaxies shown in Fig. \ref{fig:fig_denclue_spiral_arms}, for the four different combinations 
		of $h$ and $\varepsilon_d$.
		The data points are plotted as  as vertical sticks on the x axis.
		The solid lines represent the density function of the data points, estimated using a Gaussian kernel.
		\textit{Left panels:} radial distribution of  core (orange) and non core (green) \textit{density attractors} for the
		elliptical galaxy case. \textit{Right panels}: the same as in the left panels, for the spiral galaxy case. }
	\label{fig:attr_rad_distr}
\end{figure*}

\subsubsection{Morphological Features}
\label{sec:Morph_features}
 
We  extract,  from  the source cluster image, also a set of  morphological features already presented in the literature.
A detailed description of these features is given in  \cite{Conselice:2003,Lotz:2004,Conselice:2008de}. 
In the  following we give only a brief description. An important point, that differs 
from previous application of these features, is that in our case 
we do not need to deal with the background. Indeed our cluster image stores  only the pixels belonging to the  detected source cluster, hence all the background pixels are 
already removed from the image. 

\begin{itemize}[leftmargin=0.5cm]
\item \texttt{morph\_gini} - the Gini factor. The Gini factor \citep{Lotz:2004} indicates the distribution of the light among pixels.
A value of Gini equal to 1 would indicate that  all the light is concentrated in one pixel, on the contrary,
  a lower value would indicates that the light is distributed more evenly amongst the pixels, with Gini factor equal 
 to 0, if all the pixels have the same flux value. 
\item \texttt{morph\_M20} - the normalized second-order moment of the 20\% brightest pixels of the galaxy: this feature has been introduced by \cite{Lotz:2004}, and it is computed starting from the total
second order moment
 \begin{equation}
 M_{tot}=\sum_i^n M_i=\sum_i^n I_i[(x_i-x_c)^2+(y_i+y_c)^2],
 \end{equation}
  where 
$x_c,y_c$, represent the source cluster centroid, and $I_i$, is the flux of each source cluster pixel,
with coordinate $x_i,y_i$.
$M_{20}$ is defined as  the normalized second-order moment of the 20\% brightest pixels of the galaxy  computed as:
\begin{equation}
	M_{20}=	\log10 \frac{\sum_i M_i}{M_{tot}}, \textnormal{While}  \sum_i I_i<0.2I_{tot}
\end{equation}
 
\item \texttt{morph\_conc\_1,morph\_conc\_2 }  the concentration indices, defined as:
\begin{eqnarray}
C1&=&\log(r_{80}/r_{20}),\\
C2&=&\log(r_{90}/r_{50})\nonumber
\end{eqnarray}
where $r_{x}$ represents the radius of the circular discs containing  a x\% and 
of the total cluster  flux.

\item \texttt{morph\_r\_80\_to\_r\_max, morph\_r\_20\_to\_r\_max}- the ratio of  
 $r_{80}$ to $r_{max}$, and $r_{20}$ to $r_{max}$

\item \texttt{morph\_r\_Petrosian\_to\_r\_max} -  the ratio of the Petrosian radius 
to the deprojected source cluster $r_{max}$. To determine the Petrosian radius we use the same definition as reported in \cite{Conselice:2008de}. Let $\mu(r)$ be 
the surface brightness at the radius $r$, and $\mu(<r)$ the integrated surface brightness within $r$.
We define $r_{\eta}$ as the radius at which $\mu(r_{\eta})/\mu(<r_{\eta})=\eta$, and the Petrosian 
radius as $R_{p}=1.5r_{\eta=0.2}$

\item  \texttt{morph\_clumpiness}  - the clumpiness index, defined as:
\begin{equation}
S=10(\frac{\sum_i^n |I_i - I_i^*| }{\sum_i^n | I_i|}\big)
\end{equation}
where $I^*$ is the cluster image smoothed with a Gaussian Kernel, with 
a bandwidth $w=s*R_p$, where $s$ is a factor ranging in $0.0-1.0$.
\item \texttt{morph\_asymm} - the asymmetry index, defined as:
\begin{equation}
A=min\big(\frac{\sum_i^n|I_i-I_{180,i}|}{\sum_i^n| I_i|}\big)
\end{equation}
where $I_{180}$ is the cluster image rotated  180 degrees around the cluster centroid.
\end{itemize}

\begin{figure*}
	\centering
	\begin{tabular}{ll}
		\includegraphics[width=8.0cm]{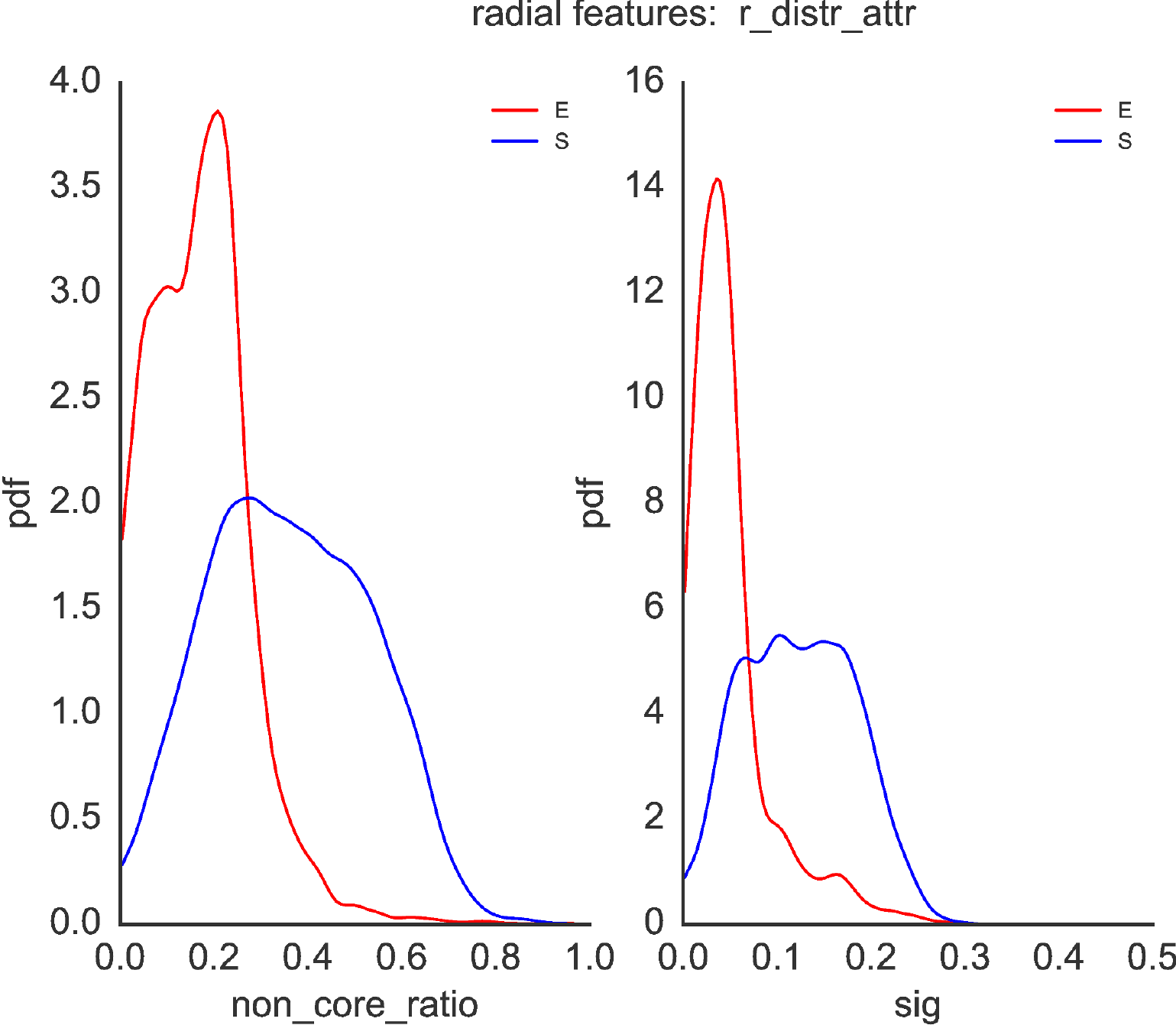} &
		\includegraphics[width=8.0 cm]{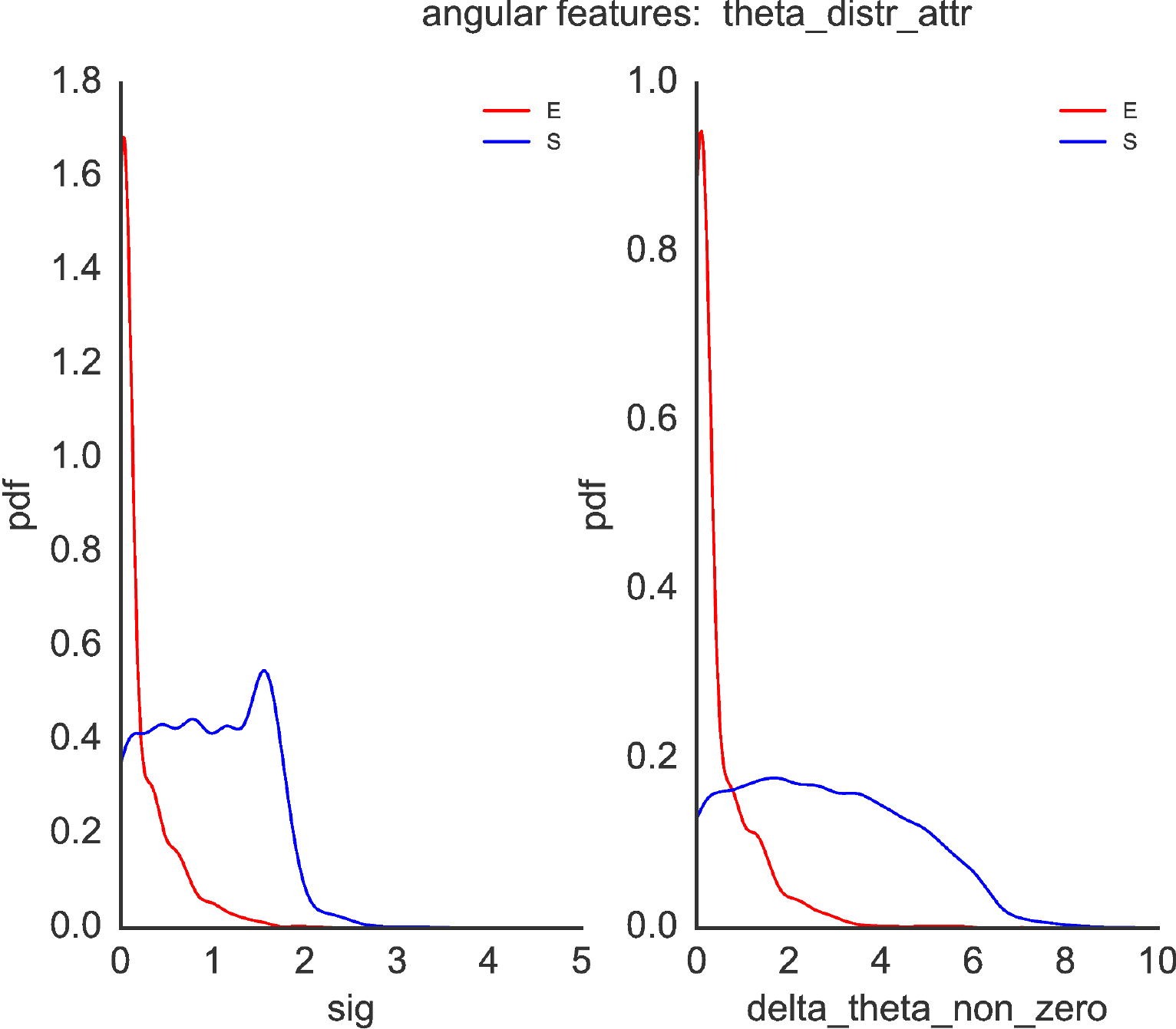} \\
	\end{tabular}
	\caption{Distribution of features derived from radial and angular distribution of non-core density 
		attractors, showing the separation between the elliptical and spiral classes, 
		 for  the objects in our sample, averaged for  four different values of the DENCLUE parameter $h$ = [1.0,1.2,1.4,1.6].
		\textit{Left panels}: distribution of \texttt{r\_distr\_attr\_non\_core\_ratio}
		and \texttt{r\_distr\_att\_delta\_r\_non\_zero}.
		\textit{Right panels}: distribution of \texttt{theta\_distr\_attr\_sig} 
		and \texttt{theta\_distr\_attr\_delta\_theta\_non\_zero}.
		The solid lines represent the density function, estimated using a Gaussian kernel.}
	
	\label{fig:attr_rad_distr_feat}
\end{figure*}

\subsubsection{Features from radial distribution of the density attractors}
\label{sec:attractors_features}
In general, elliptical galaxies have only one cluster of attractors 
corresponding to the core of the galaxy, and zero or small attractor clusters 
outside. On the contrary, spiral galaxies in addition to the core cluster, have  clusters of attractors that 
tracks the spiral arms, this reflects in a strong difference in the radial and angular distribution of 
the attractors. A clear example is given in Fig. \ref{fig:attr_rad_distr}, where we show the radial distribution of \textit{the density attractors} for the detections 
of the two galaxies shown in Fig. \ref{fig:fig_denclue_spiral_arms}. The radial distribution of the non-core 
attractors (green lines) is quite different in the two cases of elliptical and spiral objects. 
The former has zero, or very little contribution, the latter shows a significant contribution. 
In particular we note that in the case of elliptical, when non-core \textit{density attractors} are found, 
they are distributed  with a narrow peak, while elliptical have a broader distribution. 
To make the comparison more homogeneous, we normalize the  radial coordinate as $r=r/r_{max}$, 
where $r_{max}$ is the most distant point in the source cluster, from its centroid. 
Hence, we can use a statistical characterization of these distributions in order to provide important classification features. In the following we list and explain the extracted features.
The root name of the radial distribution  variables is 	\texttt{r\_distr\_attr\_}, and that for the angular distribution  variable is \texttt{theta\_distr\_attr\_}, and in the following of this section we  report  only the specific name.
We remind that all the features extracted from the radial  and angular distribution of the  \textit{density attractors},   are evaluated from the distribution of only non-core \textit{density attractors}:
The features evaluated for the radial distribution are:
\begin{itemize}[leftmargin=0.5cm]
	\item	\texttt{non\_core\_ratio} - the ratio  of number non-core \textit{density attractors} to  the total number of points 
	in the source cluster defined as:
	\begin{equation}
	\texttt{non\_core\_ratio}= \frac{N_{non-core}}{N_{source}}
	\end{equation} 

	where $N_{non-core}$ is the total number of 
	points in the non-core clusters, and $N_{source}$ is the total number of points in the source cluster.

	\item \texttt{mode} - the mode of the  radial distribution.  
	\item \texttt{sig} - the standard deviation of  the radial distribution.  
	\item \texttt{skew} - the skewness of  the radial distribution.	
	\item \texttt{skew\_p\_val} -  the two-sided p-value for the skewness test.
	\item \texttt{kurt} - the kurtosis of  the radial distribution.	

	\item \texttt{delta\_r\_non\_zero} - the difference.
	between the  largest and the smallest value of $r$, where the distribution is larger than zero.
\end{itemize}
The features evaluated for the angular distribution are:
\begin{itemize}[leftmargin=0.5cm]
	
	\item \texttt{sig} - the standard deviation of  the angular distribution.  
	\item \texttt{skew} - the skewness of  the angular distribution.	
	\item \texttt{skew\_p\_val} -  the two-sided p-value for the skewness test.
	\item \texttt{kurt} - the kurtosis of  the angular distribution.	
	
	\item \texttt{delta\_theta\_non\_zero} - the difference.
	between the  largest and the smallest value of $theta$, where the distribution is larger than zero.
\end{itemize}
Moreover, we extract also the Hu moments for the density attractors both in Cartesian and polar coordinate,  \texttt{attr\_log\_Hu\_[0-6]} and  \texttt{attr\_polar\_log\_Hu\_[0-6]}  respectively.

Left panels of Fig. \ref{fig:attr_rad_distr_feat} show the distribution of  the radial features	\texttt{non\_core\_ratio}  and 
 \texttt{sig},  for  the objects in our sample, 	averaged for  four different values of the DENCLUE parameter $h$ = [1.0,1.2,1.4,1.6]. Both spiral and elliptical objects  have the mode of the distribution  at $\texttt{non\_core\_ratio}\simeq 0.2$, but  elliptical galaxies,  as expected,   are characterized by  lower values of 	\texttt{non\_core\_ratio}, indeed $\simeq 70\%$ of the elliptical  have a value of $\texttt{non\_core\_ratio}$ below the mode, while  $\simeq 80\%$ of the spiral 
 have a value of $\texttt{non\_core\_ratio}$ above the mode.
 A further difference is given by width of the radial distribution, indeed  we observe that elliptical objects are characterized by values of standard deviation of the radial distribution, \texttt{sig}, close to zero, while spiral ones have larger values. Again, this behavior is in agreement with our  expectations, indeed, spiral arms will have \textit{density attractors} that typically will span 
the largest fraction of the radial extension of the galaxy, while \textit{density attractors} 
in elliptical will be related mostly to noisy pixels,  or to unresolved sources (i.e. cases 
in which the deblending did not succeed.), or to rings in polar ring galaxies or lensing signals,
and will have a narrower radial extent.
This effect has an impact on the angular size of the \textit{density attractors} and it is confirmed by the plots in the right  panels of  Fig. \ref{fig:attr_rad_distr_feat}, where we show the distributions of the angular features   \texttt{sig} and    \texttt{delta\_theta\_non\_zero}. Both the plots show how in the case of elliptical objects the 
  angular size of the \textit{density attractors}  peaks close to 0, while in the case of elliptical objects
  it reaches a much larger angular size.

\subsection{Final features set}
\label{sec:final_fetures_set}
In  total we have 105 features ($N_f$) extracted by our pipeline. We have run the pipeline with four different
values of the DENCLUE parameter $h=[1.0,1.2,1.4,1.6]$ . This parameter, as already discussed, has a important impact 
on the scale of the morphological structures that will be found by the attractors. The other parameters used 
for the feature extraction are reported in Tab.  \ref{tab:pipeline_par_space}.
Fore each feature, we compute the average of the four values obtained with the different h values.
We will refer to this set as the \textit{h-averaged} features set, and it will be used in the
classification analysis presented in the next section. A fits table version of  this features set is available as supplementary material (see App. \ref{App:data})

\begin{table}
	
	\centering
		\caption{Pipeline parameter space}
		\label{tab:pipeline_par_space}
	\begin{tabular}{ccc}
		\hline
	     Pipeline task & parameter & values  \\
		\hline
		\multirow{ 1}{*}{background est.}   & $N_{bkg}$ &  3.5  \\
		 \hline
		\multirow{ 1}{*}{dbscan detection}   & $K_{th}$ &  1.5  \\
		\hline
	
		\multirow{ 2}{*}{denclue deblending}  & $h$ & 0.15$\times r_{max}$ \\		
																 & $\varepsilon_d$ & 0.01\\		
		\hline
		\multirow{ 2}{*}{density attractors}   & $h$ & 1.00,1.20,1.40,1.60\ \\		
		                                                       & $\varepsilon_d$ & 0.10\\
		\hline
		\multirow{ 2}{*}{Unsharp Filtering}   &  $u$ & 0.5\\		
						                                        &  $\sigma_s$ & 1.0 $\times R_{p}$\\
		\hline	
	\end{tabular}

\end{table}

\section{Training Sets}
\label{sec:trainign_sets}
We decided to test our classification against the class labels reported in the \texttt{gz2class}  column 
of the GZ2 catalog, and against the labels extracted from the answers  to tasks \texttt{ t01} and \texttt{ t04}
of the GZ2 decision tree (see Tab. \ref{tab:gz2_tasks}).
We  have prepared three sets of training objects (see Tab. \ref{tab:labels_sets}):

\begin{itemize}[leftmargin=0.5cm]
	\item The  \textit{GZ2 class} set of training objects  is based on the class extracted from the column \texttt{gz2class}  in the GZ2 catalog, and the features from the  \textit{h-averaged} 
	 set. All objects with the \texttt{gz2class}  string starting with `E' 
	have been labeled as elliptical,  and all the objects with string starting with `S' have been labeled as spiral.

	\item The two sets of training objects,   \textit{GZ2 task 01} and \textit{GZ2 task 04}, are based on the  labels  extracted from the answer to tasks \texttt{ t01} and \texttt{ t04}. Also in this case we use the features from the  \textit{h-averaged} features set. We note that in our selected sample of objects there are no entry for the $A1.3$ answer, corresponding to the start or artifact category.

\end{itemize}
A fits table version of  these training features set is available as supplementary material (see \ref{App:data})

\section{Supervised ensemble classification}
\label{sec:superv_class}
 Supervised machine learning (ML) represents a powerful tool to infer a classification function 
 from a labeled training data set. 
 One of the possible method used is given by the so called "Ensemble"
 methods, that rely  on the combination of several  learners.
 The outcome from the combination of several  learners will be much more robust 
 than  that from a single one. Among the `Ensemble' family, there is a further  separation 
 based on the relation among the estimators in the ensemble, and their impact on the 
 classification error:
 \begin{itemize}[leftmargin=0.5cm]
 	\item \textbf{averaging based methods}: based on  large number of complex, and mutually 
 	independent estimators. Since each estimator is `complex', the error bias is low, but the variance 
 	can be large, and it is reduced by the averaging. 
 	\item \textbf{boosting based methods}: based on sequential construction of  simple estimators. 
 	The estimators are `weak', hence they have large bias, but a small variance. The combination of the 
 	estimators leads to a decrease in the bias.  	
 \end{itemize}
 
 We use two different supervised ensemble classification algorithms, Random Forest  \citep{Breiman:2003}  belonging  
 to the family of  \textit{averaging based methods}, and Gradient Boosting   \cite{Friedman:2001ic} belonging to the family 
 of  \textit{boosting based methods}. For both the classifier we have used  the 
 implementation provided   by the scikit-learn \citep{scikit-learn} python package

 \begin{figure}
 	\centering 
 	\includegraphics[width=8.0cm]{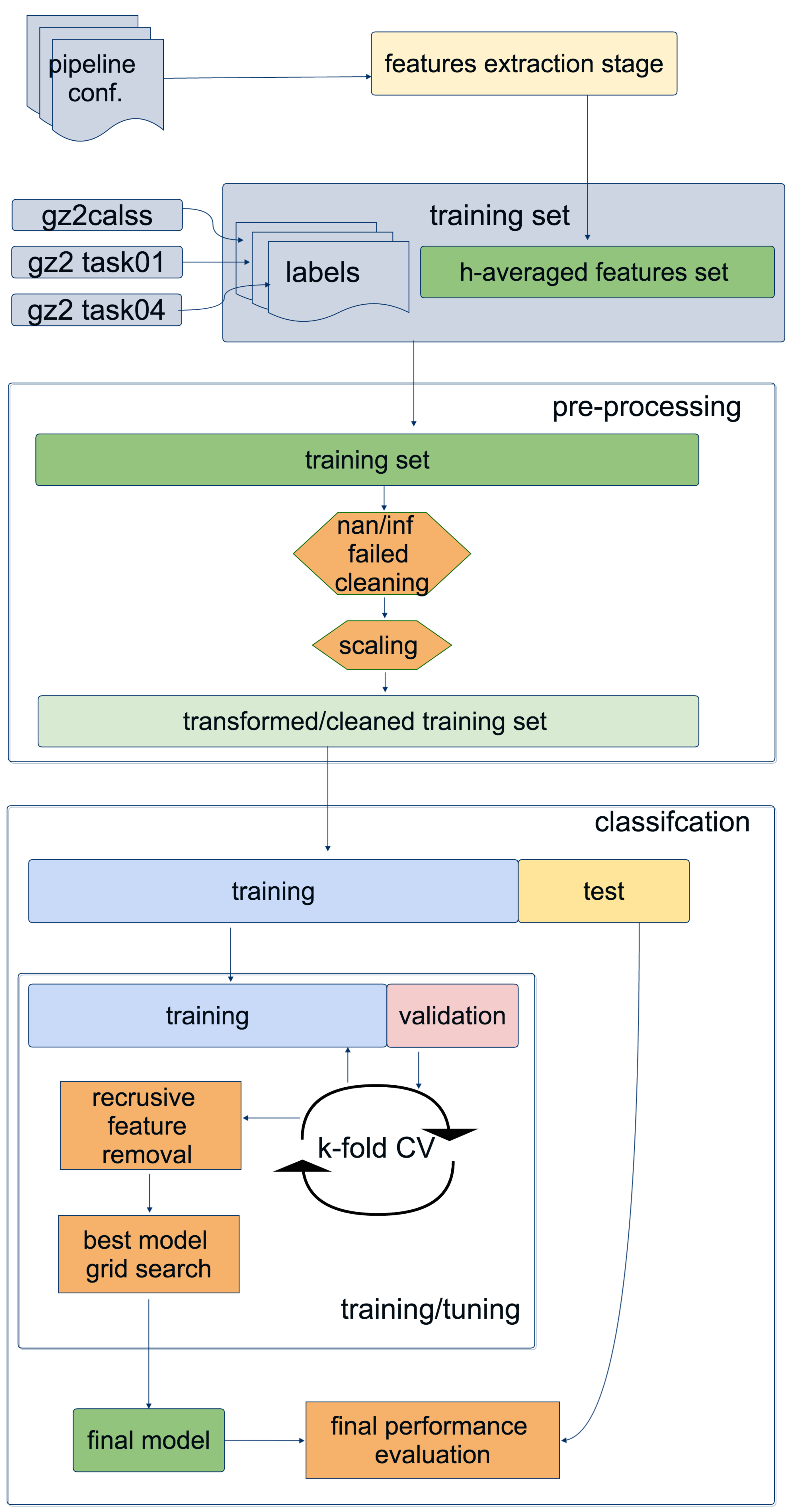}
 	\caption{Diagram for the supervised classification process.}
 	\label{fig:ml_strategy}
 \end{figure}
 
 \begin{table}

 	\caption {Sets of training objects}	\label{tab:labels_sets}
 	\centering
 	\begin{tabular}{|c|c|c|c|c|}
 		\hline
 		Name & class labels & size \\
 		\hline
 		\textit{GZ2 classes} 	               & E,S                           & 24635 \\
 		\hline
 		\textit{GZ2 task 01} 	               & A1.1,A1.2         & 24635 \\
 		\hline
 		\textit{GZ2 task 04} 	               & A4.1,A4.2                  & 21730 \\
 		\hline
 	\end{tabular}
 	
 \end{table}
 
 \subsubsection{Gradient Tree Boosting}
 
 The Gradient Tree Boosting (GB) method, 
 is based on the idea of building a strong learner from the combination of several  
 weak learners. Weak learners are decision trees, 
 and are added together sequentially. At each stage of the process, a further decision tree is added to the 
 ensemble, and the newly generated decision tree is trained in order to minimize a loss function.  
 This result is achieved by making the new decision tree maximally correlated with the negative 
 gradient of the loss function.
 A detailed review of GB methods is given in \cite{Natekin:2013ew}

 \subsubsection{Random Forest}
 Random Forest (RF)   is an ensemble classification method, based on averaging. It  has been successfully 
 applied to different astrophysical subjects, such as classification of periodicity in variable stars \citep{Dubath:2011kn}.
 Trees are constructed from bootstrapped samples of the original training set. A  randomize set has the same size than the original training set, but some 
 of the elements of the original training set   may appear multiple times, while others are missing.
 The original elements missing in the  bootstrapped sample will constitute  the so called \textit{out-of-bag} (OBB) sample.
 
 If our training set has $N_{f}$ features, each  decision tree will be built from a random subsample of features  whose number is $n_{f}<<N_{f}$. Trees are grown deep, hence will have low bias but a large variance. The averaging over a large number $M$  of trees, has the goal to reduce the variance of the final model. 
 
\subsection{classification metrics} 
To express the capacity of the model to predict correctly a given morphology, we need to provide 
a metric of the classification results. We will use the following indicators:
\begin{itemize}[leftmargin=0.5cm]
	\item \textit{accuracy}:  i.e.,  the ratio of the total number of correctly classified objects, to the total number of objects in the sample
	\item \textit{precision} or \textit{purity}, defined as  $P=\frac{TP}{TP+FP} $, where $TP$ are the true positive, and $FP$ are the 
	false positive.  This indicator gives the fraction of correctly classified objects, for a specific 
	class.
	\item \textit{recall} or \textit{completeness}, defined as  $R=\frac{TP}{TP+FN}$, where  $FN$ are the 
	false negative. This indicator gives the fraction of 
	correct prediction for a specific class, in terms of the total number of object actually belonging to that class.  
\end{itemize}

 \section{Classification strategy}
 \label{sec:class_strategy}
 
Even though ML classification methods are quite powerful, their application  requires well defined 
 strategies  that allows to fit  the `model'  to a set of training data,  with  reliable predictions on general data (never used in the training process) and without overreacting to the noise  present in the training set.  To accomplish  this goal we need to tune the model in such a way  that it is not too simple, i.e. suffering from underfitting the data (high bias), or too complex hence suffering from overfitting the data (high variance). 
Our ML strategy is based on two stages, a pre-processing stage, and a classification stage, and the corresponding pipeline has been implemented around te scikit-learn framework \citep{scikit-learn}. 
A schematic view is shown in Fig. \ref{fig:ml_strategy}.

\subsection{data pre-processing}
\label{sec:data_pre_proc}
 \begin{itemize}[leftmargin=0.5cm]
 \item \textbf{nan/inf, failde, cleaning} We remove all the entries with features having non-valid  values, or the entries corresponding to failed featured extraction.
 \item  \textbf{scaling} We scale the features in our features set using the procedure of the standardization (\texttt{StandardScaler} method from  scikit-learn ), i.e. the features are centered at mean 0, with a standard deviation of 1. Even though  this step  is not mandatory 
 for the ensamble classification methods that we will use, it  allows a better comparison with other ML models.\end{itemize} 
 
 \subsection{classification}
 \label{sec:data_classif}
 \begin{itemize}[leftmargin=0.5cm]
 \item  \textbf{train/test partitioning} the training set is split in training and a test set, with a ratio of $80\%$ training, $20\%$ test.
 The data in the test set will never be seen by the model training process, and will give our final benchmark for the model performance
\item  \textbf{k-fold cv  for model tuning} We split our training set in training and validation set, using a  k-fold cross validation 
method (\texttt{StratifiedKFold} method from  scikit-learn) that splits randomly the training set into k folds without replacement, where k-1 are used for the training and one for the validation. In particular we use a stratified k-fold cross validation, which preserves in
each fold the original relative ratios of the classes.
\item  \textbf{model tuning: dimensionality reduction} We use a recursive feature elimination based on the accuracy ranking from the cross-validate score on the model, using the the \texttt{RFECV} method   scikit-learn.
\item \textbf{model tuning: regularization} We regularize the model by  optimizing the number of decision trees, using the accuracy score from a   cross-validated grid-search  based on the \texttt{GridSearchCV}  method from scikit-learn.
\item \textbf{final performance} the final model is tested on the test stet, whose data have never been used in any of the training steps
before.\end{itemize}

 \begin{table*}
 	
 	\centering
 	\caption{Classification metrics for the   \textit{GZ2 class}  test set:
 		statistics come from the  trials corresponding to 100 repetitions of the classification experiment. The firs column gives the name of the classifier. The second column reports the accuracy. 
 		The third columns shows the morphological classes. The fifth and sixth, 
 		report the precision and the recall, respectively. The las column reports the average
 		number of features, used in the classification, after the feature selection}
 	\label{tab:gz2class_rep}
 	
 	\begin{tabular}{|C{1cm}|c|c|c|c|c|c|}
 		\hline
 		classifier           & accuracy                                                                               & Class & precision         & recall            & $<N_f>$       &Test size        \\ \hline
 		\multirow{2}{*}{RF} & \multirow{2}{*}{\begin{tabular}[c]{@{}c@{}}$0.926 \pm 0.003$ max=0.933\end{tabular}} & E     & $0.889 \pm 0.008$ & $0.853 \pm 0.008$ & \multirow{2}{*}{43.89} &\multirow{4}{*}{4927}\\
 		&                                                                                        & S     & $0.941 \pm 0.003$ & $0.956 \pm 0.004$ &           &            \\  \cline{1-6} 
 		
 		\multirow{2}{*}{GB} & \multirow{2}{*}{\begin{tabular}[c]{@{}c@{}}$0.935 \pm 0.003$  max=0.946\end{tabular}} &        E     & $0.898 \pm 0.008$ & $0.878 \pm 0.008$ & \multirow{2}{*}{52.85}& \\ 
 		&     & S     & $0.950 \pm 0.003$ & $0.959 \pm 0.004$ &                    &   \\ \hline
 	\end{tabular}
 	
 \end{table*}
 
 \begin{figure*}
 	\begin{tabular}{c}
 		\includegraphics[width=15.0cm]{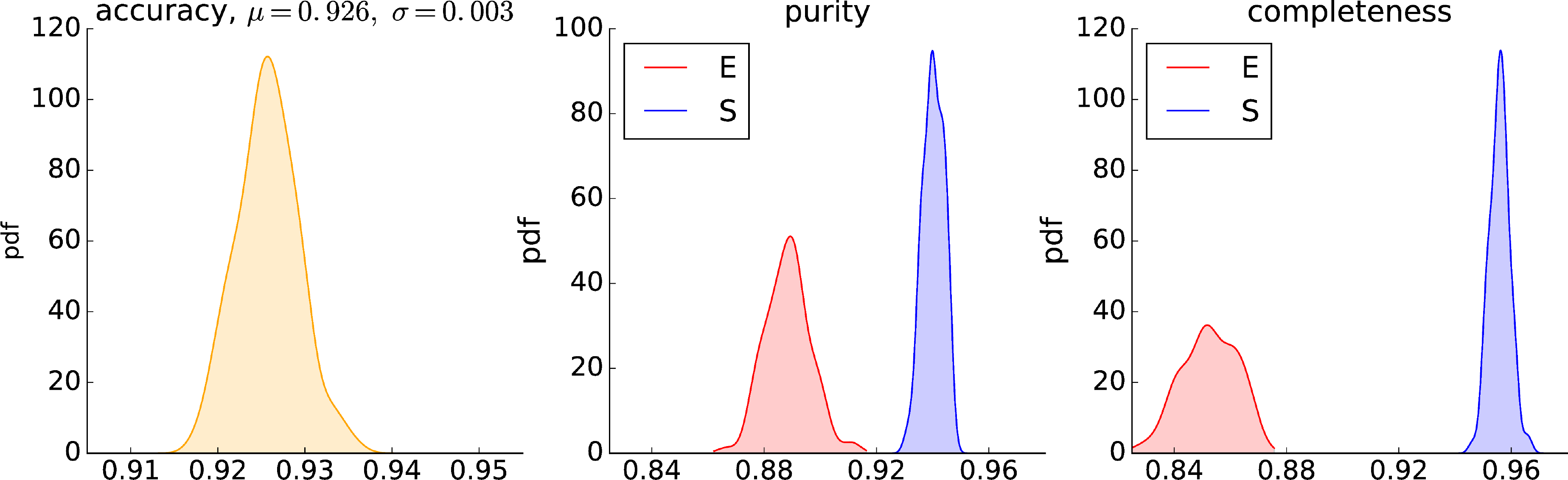}\\
 		\\
 		\includegraphics[width=15.0cm]{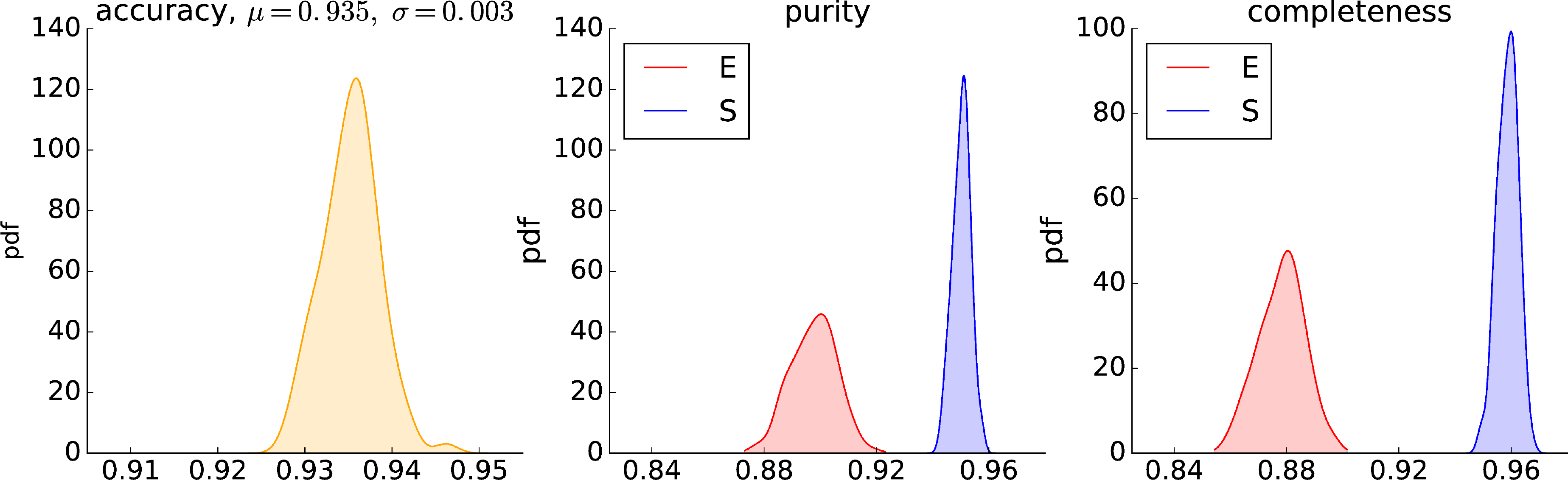}\\
 	\end{tabular}
 	\caption{
 		distribution of accuracy, purity, and completeness, for the RF (upper panels) and GB (bottom panels) classifier, for the  \textit{GZ2 class} test set
 		trials, corresponding to the  100  repetitions	 of the classification experiment.}
 	\label{fig:class_rep}
 \end{figure*}

\section{Classification performance}
\label{sec:class_performance}
 In order to assess the classification performance of our method we repeat 100 times a classification experiment, 
 using the methodology describe in \ref{sec:data_classif}, and illustrated in the classification box of Fig. \ref{fig:ml_strategy}. 
 Before running the classification we pre-process the data as reported described in \ref{sec:data_pre_proc}. 
 The cleaning process removes only two entries from our training set, hence we have final number of 24633 
 objects in the \textit{gz2class} training set.

  \begin{figure*}
  	\begin{tabular}{ll}
  		\includegraphics[width=8.5cm]{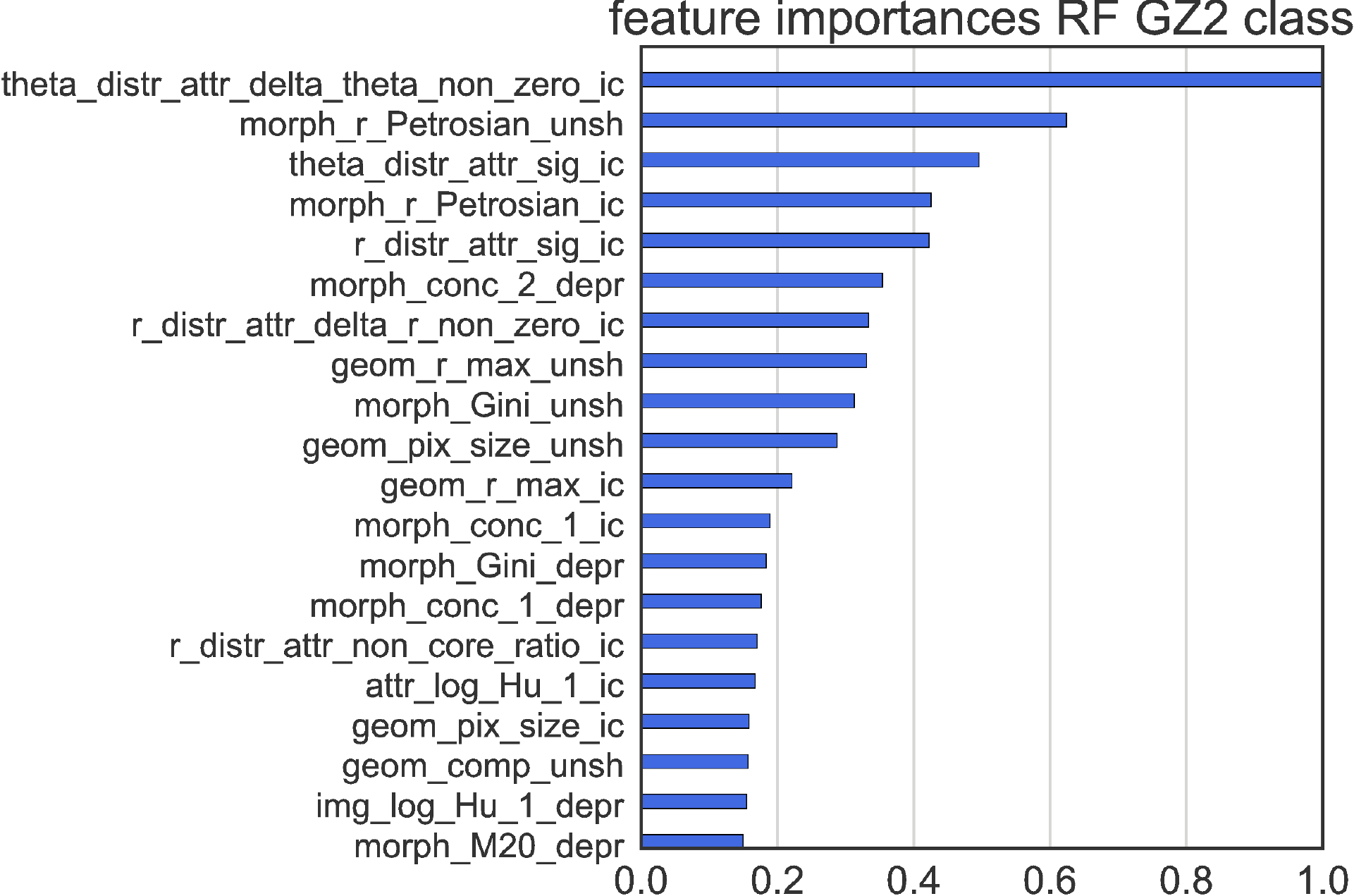}
  		&\includegraphics[width=8.5cm]{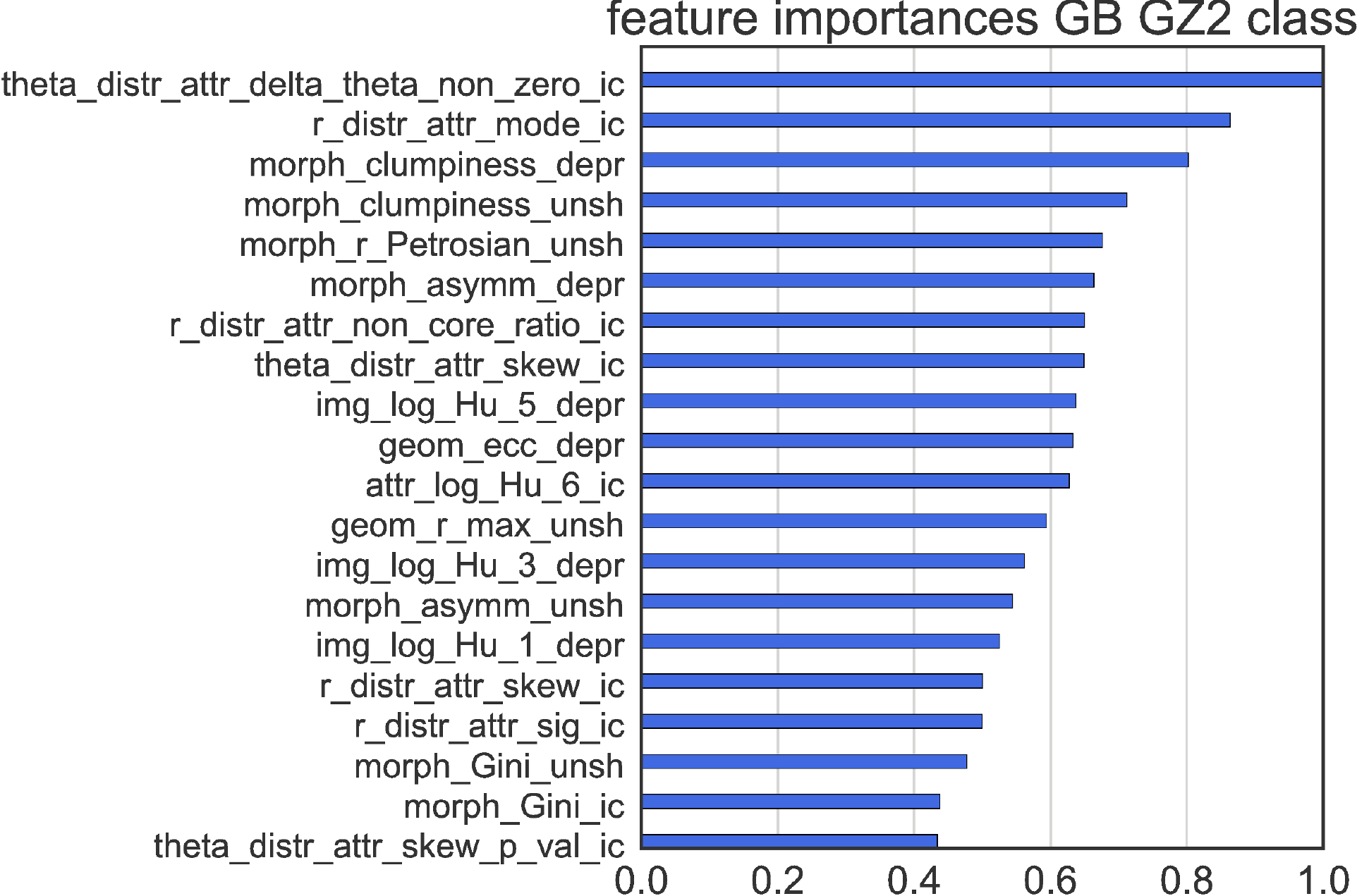}\\
  	\end{tabular}
  	\caption{Top 20 ranked features importances, for RF (left panel) and GB (right panel), classifiers, averaged 
  		over the 100 classification experiment repetitions with the  \textit{GZ2 class} test set.}
  	\label{fig:feat_imp}
  \end{figure*}
 
  \begin{table*}
  	
  	\centering
  	\caption{Classification metrics for \textit{GZ2 task} for the task 01 and the task 04 test sets, for the RF classifier.
  		Statistics come from the  trials corresponding to 100 repetitions of the classification experiment. Column description  as in Tab. \ref{tab:gz2class_rep}}
  	\label{tab:gztask1_rep}
  	
  	\begin{tabular}{|c|c|c|c|c|c|c|c|}
  		\hline
  		classifier          & GZ2 Task             & accuracy                                     & Answer & precision         & recall            & $<N_f>$     & Test set size          \\ \hline
  		\multirow{5}{*}{RF} & \multirow{3}{*}{t01} & \multirow{3}{*}{$0.922 \pm 0.004$ max=0.931} & A1.1   & $0.887 \pm 0.008$ & $0.847 \pm 0.009$ & \multirow{3}{*}{43.2} &\multirow{3}{*}{4927}\\ 
  		&                      &                                              & A1.2   & $0.936 \pm 0.004$ & $0.954 \pm 0.004$ &                     &  \\
  		&                      &                                              & A1.3   & - & - &                       &\\ \cline{2-7} \cline{7-8} 
  		& \multirow{2}{*}{t04} & \multirow{2}{*}{$0.929 \pm 0.003$ max=0.936} & A4.1   & $0.949 \pm 0.003$ & $0.964 \pm 0.003$ & \multirow{2}{*}{39.9}&\multirow{2}{*}{4346} \\  
  		&                      &                                              & A4.2   & $0.82 \pm 0.01$ & $0.77 \pm 0.01$ &                      & \\ \hline
  		
  	\end{tabular}

  	\centering
  	\caption{Same as in Tab. \ref{tab:gz2class_rep} for the GB classifier.}
  	\label{tab:gztask4_rep}
  	
  	\begin{tabular}{|c|c|c|c|c|c|c|c|}
  		\hline
  		classifier          & GZ2 Task             & accuracy                                     & Answer & precision         & recall            & $<N_f>$     & Test set size          \\ \hline
  		\multirow{5}{*}{GB} & \multirow{3}{*}{t01} & \multirow{3}{*}{$0.930 \pm 0.003$ max=0.936} & A1.1   & $0.892 \pm 0.007$ & $0.869 \pm 0.009$ & \multirow{3}{*}{52.3} &\multirow{3}{*}{4927}\\  
  		&                      &                                              & A1.2   & $0.945 \pm 0.004$ & $0.955 \pm 0.003$  &                      & \\ 
  		&                      &                                              & A1.3   & - & - &                      & \\ \cline{2-7}  \cline{7-8} 
  		& \multirow{2}{*}{t04} & \multirow{2}{*}{$0.931 \pm 0.003$ max=0.940} & A4.1   & $0.954 \pm 0.003$ & $0.963 \pm 0.003$ & \multirow{2}{*}{58.6} &\multirow{2}{*}{4346}\\
  		&                      &                                              & A4.2   & $0.82 \pm 0.01$ & $0.79 \pm 0.02$ &                       &\\ \hline
  	\end{tabular}
  	
  \end{table*}
 
 \subsection{GZ2 classes training set}
 The train/test partitioning is done with a ratio of $80\%$ training  $20\%$ test, for  a corresponding 
 number of 19706 objects for the training set, and 4927 for the test set.
 The results for the test set classification are reported in Tab. \ref{tab:gz2class_rep}, and shown in Fig. \ref{fig:class_rep}. The average accuracy of the classification is $\simeq 93\%$ for the RF classifier 
 and $\simeq 94\%$ for the GB classifier. Precision and recall for both the elliptical and spiral 
 classes are also reported.
 The overall performance is very good if we consider that the use a quite large test set of $~5000$ objects, 
 and that by repeating the experiment 100 times we give also a good estimate of the error intrinsic to the model and to the data.
 It is interesting to understand which feature have the highest classification power. We rank the features importance  using the \texttt{feature\_importances\_} attribute, that is implemented in all the tree-based models of scikit-learn. The  \texttt{feature\_importances\_}  is evaluated according to the so called `Gini importance' or `mean decrease impurity' \citep{Breiman84}. This method  defines the importance  as the total decrease in node impurity ,weighted by the probability of reaching that node, and averaged over all trees of the ensemble.
 The results are reported in Fig. \ref{fig:feat_imp}, where we plot the 20 top-ranked features for both the RF (left panel)  and the GB (right panel) classifier. We report the importance  averaged over the outcomes  of the 100 repetitions of our classification experiment.   We note that both in the case of RF and GB classifier, the most important features is \texttt{theta\_distr\_attr\_delta\_theta\_non\_zero\_ic}, meaning that the angular distribution of the attractors is storing the largest fraction of discrimation power among all the features.  The second most important feature, in the case of the RF classifier, is the Petrosian Radius of the `unsharp' cluster, and the most important features from the radial distribution of the `density' attractors is the \texttt{r\_distr\_attr\_sig\_ic}. In the case of the GB classifier, the  most important features from the radial distribution of the `density' attractors is \texttt{r\_distr\_attr\_mode\_ic}, ranking in the second position. 
We report in Fig. \ref{fig:image_stamps1} and Fig. \ref{fig:image_stamps2}, some of the images from the test set, with their GZ2 specobjid identifier, and both the actual and predicted 
classes.

 \subsection{GZ2 tasks training sets}
 As further experiment to assess the discrimination power of our classification  technique we have tested our model  against the  labels  extracted from the answers to tasks \texttt{ t01} and \texttt{ t04}. Results are summarized in Tab. \ref{tab:gztask1_rep} and Tab. \ref{tab:gztask4_rep}. We note a very good agreement between the outcome of our classifier and the GZ2 votes, with our GB classifier showing an average accuracy 
 of $\simeq 93\%$ both for the task  \texttt{ t01}, and the task  \texttt{ t04}.

 \subsection{Comparison with similar works}
In the last  years several methods, based on the application of machine learning, have been used to classify galaxy morphology.  \cite{Banerji:2010iq} and  \cite{Dieleman2015} have
 used artificial neural networks (ANN), while  \cite{HuertasCompany2008,HuertasCompany2011}  have used   Support Vector Machine (SVM) classifier and \cite{Ferrari2015} have used 
  Linear Discriminant Analysis (LDA) classifier. 

\cite{Banerji:2010iq} have used a sample of 75000 object (50000 for training  and 25000 validation) from the Galaxy Zoo 1 catalog, to train a neural network in order 
to reproduce the human classification in early type, spiral, and point source/artifact . The authors obtained the best performance (better 
than 90\%) in terms of agreement with human classification, when adding   $(g-i)$ and $r-i$ colors, to features deriving from de Vaucoleurs and exponential  profile fitting, and quantities deriving form adaptive  moments and image texture.  Performance of \cite{Banerji:2010iq}  are comparable  with our results, with the difference that we use a larger amount of features, but we do not use color information.

 \cite{Dieleman2015}  used a
   a training set of 61578 images,  and  79975 images for the validation,  from Galaxy Zoo 2, to train their neural network to reproduce the human answers to the 
  11 tasks in Galaxy Zoo 2. The results presented by the authors    refers only to  a sub sample of the test set images 
 with at least 50 percent of participant answering. In our work we have investigated two task in common with \cite{Dieleman2015}:\texttt{ t01} and \texttt{ t04}.
\cite{Dieleman2015}  evaluated their classification metrics  in 10 different bins of participants
  agreement. For the task \texttt{ t01} , they report an agreement-averaged accuracy of $87.79\%$  for a test set of 6144 objects, 
  and an agreement-averaged accuracy of $82.52\%$ for task  \texttt{ t04}, with a test set of 2449 objects. We note that for both 
  of the tasks our classification pipeline gives an accuracy  higher than that obtained by  \citep{Dieleman2015}.

In general, compared to result from neural network, we observe that our classification technique performs very well, with the advantage 
that we can have a direct understanding of the weight of the input features on the final classification.

 Finally we compare our result to those presented by \cite{Ferrari2015} and \cite{HuertasCompany2008,HuertasCompany2011}.
 
  \cite{HuertasCompany2008} classified a test set of $\simeq 1500$ objects from SDSS, and they obtained a mean accuracy 
  of $\simeq 80\%$ for the early-vs-late type classification, using a SVM classifier, hence our classification performance seems to show
   a significantly larger accuracy. 
   In a more recent work,  \cite{HuertasCompany2011}  analyzed a large sample of  $\simeq 700$k galaxies from 	the SDSS DR7 spectroscopic sample, and they provided a probability
  of the object to belong to a given class. Since their classification scheme is based on 4 morphological types, it is not straightforward to compare \cite{HuertasCompany2011} results   to ours.
  
  Among the works found in the literature the  analysis approach presented in \cite{Ferrari2015}  is the one that is closest to ours, indeed the authors 
  use purely morphometric information (no colors), and a classifier (LDA) rather than a neural network. \cite{Ferrari2015} analyzed three 
  different data sets, the EFIGI catalog \citep{Baillard2011}, the Nair and Abraham \citep{Nair:2010bu} catalog, and the SDSS DR7 
  complete Legacy sample. Their code, Morfometryka \citep{Ferrari2015} uses as input features standard morphological coefficients, with new parameters
  such as the image entropy index $H$, and the spirality $\sigma_\phi$. This last parameter is able to compute the amount non radial 
  patterns in the image, by computing  the gradient of the polar-projected image.
 Performance of accuracy obtained by \cite{Ferrari2015} refers to the classification of a sub-samples of their catalog having a Galaxy Zoo 2
  classification, and these authors classify against the elliptical-vs-spiral  labels like in our case.
  The performance of their classification scheme, in terms of accuracy, are comparable to ours, with the difference that they use 
  a  larger data-set, but they rely only on 10-fold cross-validation, without a specific test set. Moreover, we note that in their case 
  the largest importance is obtained by features related to light concentration, unlike in our case, where the most important 
  features are related to the angular distribution of density attractors, both for the RF, and the GB classifiers.

\section{Conclusions and future developments}
 \label{sec:conclusions}
 The results presented in this work show  the successful application of \texttt{ASTErIsM}  software, based on topometric clustering algorithms (DBSCAN and DENCLUE),  to automatic galaxy detection and shape classification.
 For the detection process we have found that:
 \begin{itemize}[leftmargin=0.5cm]
 	
 	\item  DBSCAN clusters usually preserve
 	the actual shape of the source, allowing to follow quite well the contour of any 
 	arbitrary morphology.
 	
 	\item When sources are `confused', the application of the DENCLUE algorithm allows 
 	to deblend them.
 	
 \end{itemize}
 We have verified that, in addition to deblending,  the \textit{density attractors} evaluated by the DENCLUE algorithm track quite well spiral arms features, and we have found that:
 \begin{itemize}[leftmargin=0.5cm]
 	\item In general, elliptical galaxies have a single cluster of e \textit{density attractors}
 	related to the core of the galaxy, while spiral galaxies have additional ones related to 
 	the presence of spiral arms.
 	\item The  radial and angular distribution of the \textit{density attractors} are very different in the case of spiral and elliptical objects. 
 \end{itemize}
 Basing on these results we have defined a new set of features for the galaxy classification, that maximize 
 the information given by the DBSCAN clusters (see Sec. c. \ref{sec:geom_features}, \ref{sec:Hu_moments}), and the DENCLUE  \textit{density attractors} (see Sec.  \ref{sec:attractors_features}).
 In addition to these clustering-related features we have also evaluated classical  morphological  features (see 
 Sec. \ref{sec:Morph_features}).
 
 We have tested the classification performance  of the features evaluated by our pipeline, on a training set of about 24k objects, selected  from GZ2 SDSS main sample with spectroscopic redshift, using a Random Forest and a Gradient Tree Boosting  classifier. We have tested the classification performance against the GZ2 classification in elliptical vs spiral classes, and against the answers to the task  \texttt{ t01} and  \texttt{ t04} of the GZ2 decision tree. In general the accuracy of our classification, for the test set, is  $\simeq 93\%$  with a 
 performance  comparable   to other approach based on ANN \citep{Dieleman2015,Banerji:2010iq} or based on SVN and LDA classifiers \citep{HuertasCompany2008,Ferrari2015}.
  
  As future developments, we would like to investigate how deal with the classification of a larger number of morphologies, in particular investigating the capabilities to detect bars and bulges. Moreover, we aim at using the density attractors as a baseline to fit spiral arms, and investigating how this compare to human identified arms.
  We plan also to improve the DENCLUE-based deblending, using a ML approach, adding  a feedback between \textit{density attractors}  extracted  in the deblending process, and the \textit{density attractors}   extracted in the morphological feature extraction process.

\section*{Acknowledgments}
We would like to thank to anonymous referee for his/her very insightful  and useful suggestions and comments, 
which have significantly improved this work.

\appendix
\section{Technical details about the  \texttt{ASTErIsM} pipeline implementation}
\label{App:Tech_details}
The \texttt{ASTErIsM}  software is implemented as a python 2.7 package. The code is object oriented. All the algorithms for clustering and features extraction have been implemented 
from the scratch. The kernel  computation in the  DENCLUE algorithm has been 
written in Cython \citep{cython} to speed up the computational time.
Some image processing tasks are performed using the \texttt{ndimage} package from the SciPy  library  \citep{scipy},   the scikit-image \citep{scikit-image} package,
and  the Python wrapper of the  OpenCV library \citep{opencv}.
The I/O operations for the fits file use the PyFits \footnote{PyFits is a product of the Space Telescope Science Institute,	which is operated by AURA for NASA} library.  
The classification module uses the scikit-learn Python package \citep{scikit-learn}.
The graphical output, including the figure in the present paper, have been implemented 
using the Matplotlib library \citep{matplotlib} and the Seaborn \citep{seaborn} library.
 The code will be available at \url{https://github.com/andreatramacere/asterism}.
 
\section[]{Hu moments}
\label{App:Hu_moments}
The two-dimensional $(p+q)-th$ order geometric moment of a   two-dimensional 
distributions of points $(x_{i},y_{i})$  is defined as:
\begin{equation}
m_{pq}= \sum_{i=0}^{N-1} \sum_{j=0}^{N-1} (x_{j})^p (y_{j})^q
\end{equation}
If we are interested in the moments of a digital image whose pixels 
have coordinates $(x_{i},y_{i})$, and fluxes $f(x_i,y_i)$ then the previous equation reads:
\begin{equation}
m_{pq}= \sum_{i=0}^{N-1} \sum_{j=0}^{N-1} (x_{j})^p (y_{j})^q f(x_i,y_i)
\end{equation}

The centroid  can be evaluated  as:
$\bar{x}=m_{10}/m_{00}, \bar{y}=m_{01} /m_{00}$, and  the 
central moments as:
\begin{equation}
\mu_{pq}= \sum_{i=0}^{N-1} \sum_{j=0}^{N-1} (x_{j}-\bar{x})^p (y_{j}-\bar{y})^q
\end{equation}
or in the case of digital image as:
\begin{equation}
\mu_{pq}= \sum_{i=0}^{N-1} \sum_{j=0}^{N-1} (x_{j}-\bar{x})^p (y_{j}-\bar{y})^q f(x_i,y_i)
\end{equation}
The normalized central moments are given by:
\begin{equation}
\eta_{ji}=\frac{\mu_{ji}}{\mu_{00}^{(1+\frac{i+j}{2})}}
\end{equation}

As proved by \cite{Hu:1962}, it is possible to obtain moments that are invariant
under translation, scaling and rotation:
\begin{eqnarray}
Hu[0]&=&\eta_{20}+\eta_{02}\\
Hu[1]&=&(\eta_{20}+\eta_{02})^2+4\eta_{11}^2 \nonumber\ \\
Hu[2]&=&(\eta_{30}+3\eta_{12})^2+(3\eta_{21}+\eta_{03})^2 \nonumber \\
Hu[3]&=&(\eta_{30}+\eta_{12})^2+(\eta_{21}+\eta_{03})^2 \nonumber \\
Hu[4]&=&(\eta_{30}-3\eta_{12})(\eta_{30}+\eta_{12})^2 \nonumber \\
&&[ (\eta_{30}-3\eta_{12})^2 - 3(\eta_{21}+\eta_{03})^2 ] \nonumber \\
&&+(3\eta_{21}-3\eta_{03})(3\eta_{21}+\eta_{03}) \nonumber\\
&&[ 3(\eta_{30}+\eta_{12})^2 -(\eta_{21}+\eta_{03})^2 ] \nonumber \\
Hu[5]&=&(\eta_{20}-\eta_{02})[(\eta_{30}+\eta_{12})^2- (\eta_{21}+\eta_{03})^2]+ \nonumber\\
&& 4\eta_{11}(\eta_{30}+\eta_{12})(\eta_{21}+\eta_{03})\nonumber\\
Hu[6]&=&(3\eta_{21}-\eta_{03})(\eta_{21}+\eta_{03}) \nonumber \\
&&[ 3(\eta_{30}+\eta_{12})^2 - (\eta_{21}+\eta_{03})^2 ] \nonumber \\
&&-(\eta_{30}-3\eta_{12})(\eta_{21}+\eta_{03}) \nonumber\\
&&[ 3(\eta_{30}+\eta_{12})^2 -(\eta_{21}+\eta_{03})^2 ] \nonumber \\
\end{eqnarray}

\section{Data products and tables}
\label{App:data}
We provide a fits table of the training sets used in this work. The table contains the columns concerning the  \textit{h-averaged} features set,  
as well as those corresponding to \textit{gz2class} E/S classification, and  those corresponding to the
the task  \texttt{ t01} and  \texttt{ t04} of the GZ2 decision tree. A description of the table is given in Tab \ref{tab:online_data}

 \begin{table*}
 	\centering
 	\caption{Description of the columns for the online training set. For the features names we report the string for root variable name and the string for the flag corresponding to the 
 		type of cluster used to extract the features. The specific name of the variable is left as a blank space. Please, refer to \ref{sec:features_pipeline} and Tab. \ref{tab:features} a
 		specific description of the features names  }
 	\label{tab:online_data}
\begin{tabular}{l|l}

	\hline 
	Col name & Col description \\ 
	\hline 
 	specobjid &  match to the DR8 spectrum object \\
 	ra &   right ascension [J2000.0], decimal degrees\\
 	dec & declination [J2000.0], decimal degrees\\
 	gz2class &  \textit{gz2class} label for the  E/S classification\\
 	t01  & label for the  \texttt{ t01} the GZ2 decision tree\\
 	t04 & label for the  \texttt{ t04} the GZ2 decision tree\\
 	\texttt{id\_ cluster} & The id of the  \texttt{ASTErIsM}  detected cluster corresponding to the GZ2 source. \\
 	                            & If negative, it means no source was detected, or source detection failure. \\
 	\texttt{geom\_ \_ic}  &   (7 columns)  geometrical features for the `initial ' cluster(see. \ref{sec:features_pipeline}, \ref{sec:geom_features}, and Tab. \ref{tab:features}) 	\\
 	\texttt{geom\_ \_derp} &  (7 columns) geometrical features for the `deprojected' cluster (see.  \ref{sec:features_pipeline}, \ref{sec:geom_features}, and Tab. \ref{tab:features})\\
 	\texttt{geom\_ \_unsh}  &  (7 columns)  geometrical features for the `unsharp' cluster (see.  \ref{sec:features_pipeline}, \ref{sec:geom_features}, and Tab. \ref{tab:features})\\

 	\texttt{cnt\_log\_Hu \_ic}  &  (7 columns) Hu moments for the  `initial ' cluster contour  (see.  \ref{sec:features_pipeline}, \ref{sec:Hu_moments} and Tab. \ref{tab:features})\\
 	\texttt{img\_log\_H \_ic}  &  (7 columns)   Hu moments for the  `initial ' cluster image (see.  \ref{sec:features_pipeline}, \ref{sec:Hu_moments}   and Tab. \ref{tab:features})\\
 	
 	\texttt{attr\_polar\_log\_Hu \_ic}  &  (7 columns)    (see.  \ref{sec:features_pipeline}, \ref{sec:attractors_features}, and Tab. \ref{tab:features})\\
 	\texttt{attr\_log\_Hu\_ \_ic}  &  (7 columns)    (see.  \ref{sec:features_pipeline},  \ref{sec:attractors_features}, and Tab. \ref{tab:features})\\
	
	\texttt{cnt\_log\_Hu\_ \_depr}  &  (7 columns) Hu moments for the  `derprojected ' cluster contour   (see.  \ref{sec:features_pipeline},  \ref{sec:Hu_moments} , and Tab. \ref{tab:features})\\
	\texttt{img\_log\_Hu \_depr}  &  (7 columns)  Hu moments for the  `derprojected ' cluster image    (see.  \ref{sec:features_pipeline},  \ref{sec:Hu_moments} , and Tab. \ref{tab:features})\\
	
	\texttt{morph\_ \_ic}  &  (10 columns)   morphological features for the  `initial ' cluster (see.  \ref{sec:features_pipeline}, \ref{sec:Morph_features}, and Tab. \ref{tab:features})\\
	\texttt{morph\_ \_depr}  &   (10 columns)   morphological features for the  `derprojected ' cluster (see.  \ref{sec:features_pipeline}, \ref{sec:Morph_features}, and Tab. \ref{tab:features})\\
	\texttt{morph\_ \_unsh}  &   (10 columns)   morphological features for the  `unsharp ' cluster (see.  \ref{sec:features_pipeline}, \ref{sec:Morph_features}, and Tab. \ref{tab:features})\\
	\texttt{r\_distr\_attr\_ \_ic}  &  (7 columns)   radial  distribution features for the density attractors (see.  \ref{sec:features_pipeline}, \ref{sec:attractors_features}, and Tab. \ref{tab:features})\\
	\texttt{theta\_distr\_attr\_ \_ic}  &  (5 columns)     angular  distribution features for the density attractors (see.  \ref{sec:features_pipeline}, \ref{sec:attractors_features}, and Tab. \ref{tab:features})\\

	\hline 
\end{tabular} 

 \end{table*}

\begin{figure*}

		\includegraphics[width=19.3cm]{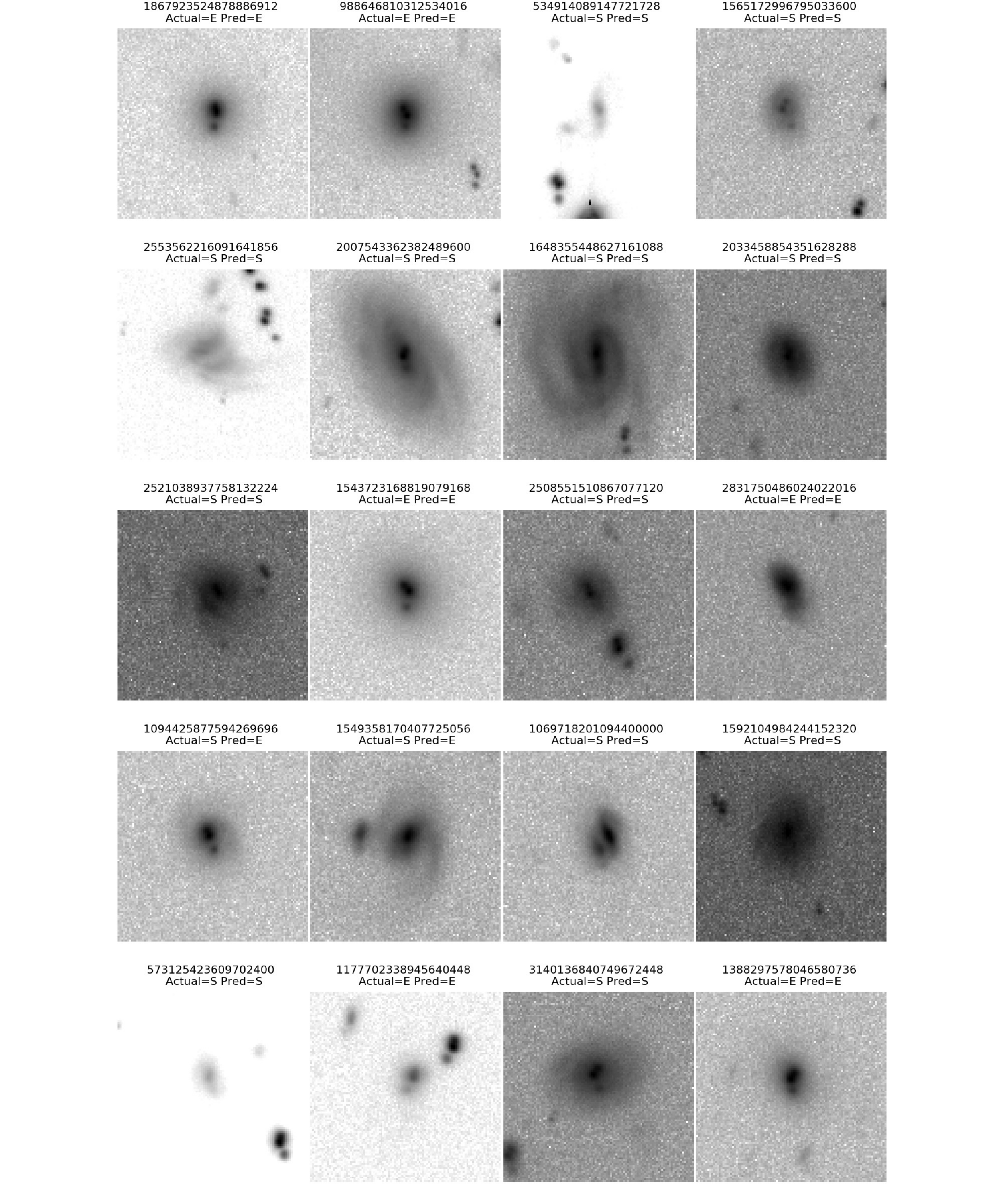}
	
	\caption{Random selection of stamps, in logarithmic flux scale, for some of the sources in the \textit{gz2class} test set. The number in the first row of the image title is the GZ2 specobjid. We report also the actual type (E/S) and the one predicted by of the runs of our GB classification experiment. }
	\label{fig:image_stamps1}
\end{figure*}

\begin{figure*}
	
	\includegraphics[width=19.3cm]{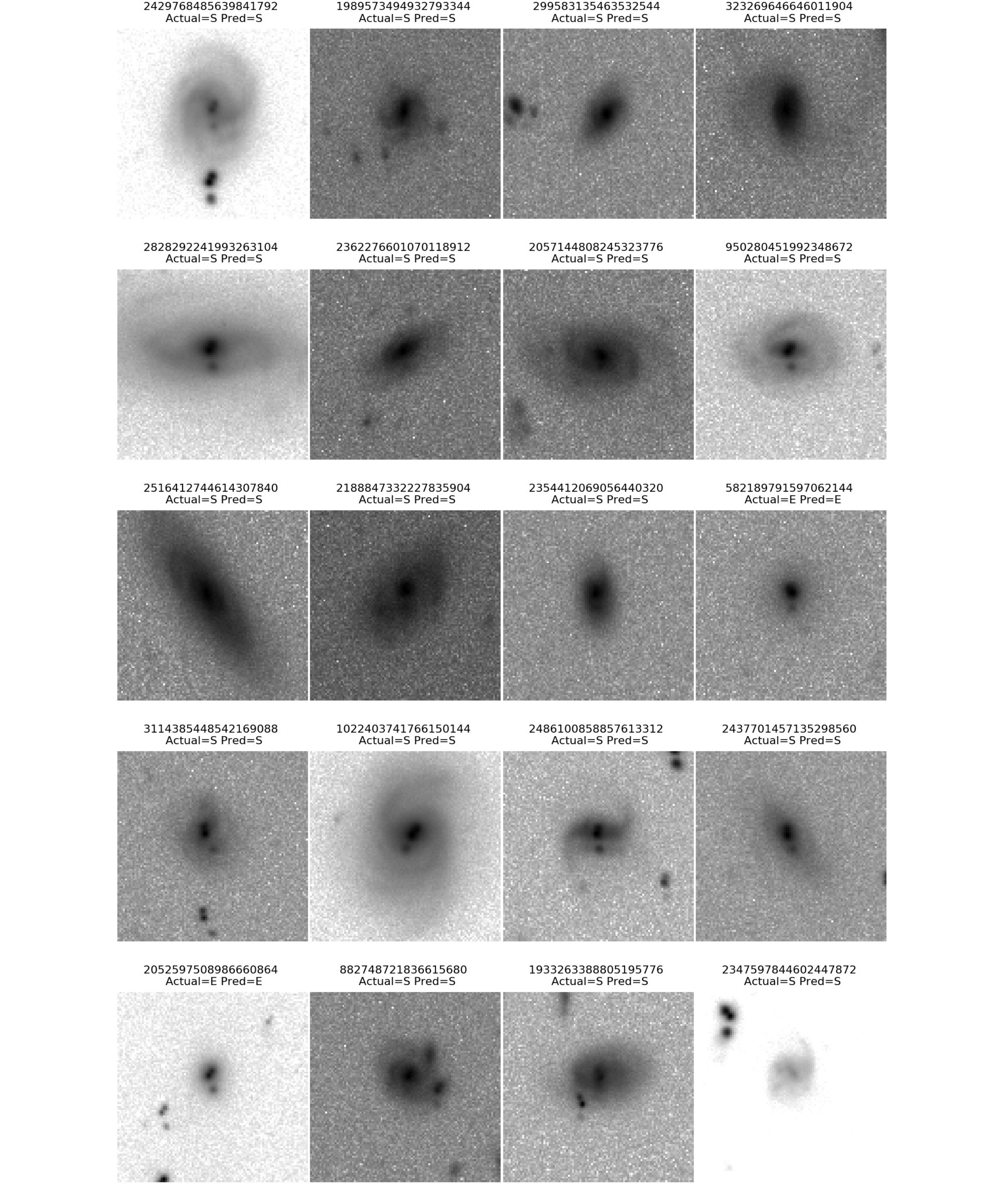}
	
	\caption{Same as for Fig. \ref{fig:image_stamps1} }
	\label{fig:image_stamps2}
\end{figure*}

\label{lastpage}
\bibliographystyle{mn2e} 
\bibliography{gal_identif}
\end{document}